\documentclass[letterpaper]{aa}
\usepackage{graphicx}
\usepackage{txfonts}
\usepackage{natbib}
\bibpunct{(}{)}{;}{a}{}{,}
\begin{document}
   \title{Near-infrared low-resolution spectroscopy\\ of Pleiades L-type brown dwarfs}

   \titlerunning{Near-infrared spectroscopy of Pleiades L-type brown dwarfs}
   \author{G.~Bihain\inst{1}\fnmsep\inst{2}
      \and R.~Rebolo\inst{1}\fnmsep\inst{2}\fnmsep\inst{3}
      \and M.~R.~Zapatero~Osorio\inst{4}
      \and V.~J.~S.~B{\'e}jar\inst{1}\fnmsep\inst{3}
      \and J.~A.~Caballero\inst{5,4}
          }

   \authorrunning{Bihain}
    \institute{Instituto de Astrof\'{\i}sica de Canarias, c/ V\'{\i}a L\'actea, s/n, 
38205  La Laguna, Tenerife, Islas Canarias, Spain\\
	     \email{gbihain@iac.es}
	     \and
	     Consejo Superior de Investigaciones Cient\'{\i}ficas (CSIC), Spain
	     \and
	     Departamento de Astrof\'{\i}sica, Universidad de La Laguna, 38205 La Laguna, Tenerife, Islas Canarias, Spain
	     \and
             Centro de Astrobiolog\'\i a (CSIC-INTA), Carretera de Ajalvir, km 4. E-28850 Torrej\'on de Ardoz, Madrid, Spain.
	     \and
             Dpto. de Astrof\'{\i}sica y Ciencias de la Atm\'osfera, Facultad de F\'{\i}sica, Universidad Complutense de Madrid, E-28040 Madrid, Spain
             }

   \date{Received 15 November 2009 / accepted 18 May 2010}
 
  \abstract 
   {The fundamental properties   of brown dwarfs evolve  with age. Models
describing  the evolution of  luminosities and effective temperatures, among
other physical parameters, can  be empirically constrained using brown dwarfs of
various masses in star clusters of well determined age  and metallicity.}
  {We aim to carry out a spectroscopic and photometric characterization of low-mass brown dwarfs  of the $\sim$120~Myr 
old \object{Pleiades} open cluster.}
  {We obtained low-resolution near-infrared spectra of the $J$ = 17.4--18.8~mag
candidate L-type brown dwarfs \object{PLIZ~28} and 35, \object{BRB~17}, 21, 23, and 29, which are Pleiades
members by photometry and proper motion. We also obtained spectra of the
well-known $J$ = 15.4--16.1~mag  late M-type cluster members \object{PPl~1}, \object{Teide~1}, and \object{Calar~3}.}
   {We find that the former six objects have early- to mid-L spectral types and
confirm previously reported M-types for the other three objects. The spectra of
the L0-type BRB~17 and PLIZ~28 present a triangular $H$-band continuum shape,
indicating that this peculiar spectral feature persists until at least the age
of the Pleiades. We add to our sample 36 reported M5--L0-type cluster members,
collecting their $I_{\rm C}$- and UKIDSS $ZYJHK$-band photometry. We confirm a
possible interleaving of the Pleiades and field L-type sequences in the $JHK$
absolute magnitude versus spectral type diagrams, and quantify marginally redder
Pleiades $J-K$ colours, by 0.11$\pm$0.20~mag, possibly related to both reddening
and youth. Using field dwarf bolometric correction -- and effective temperature
-- spectral type relations, we obtain the Hertzsprung--Russell diagram of the
Pleiades sample.  Theoretical models reproduce well the spectral sequence at
M5.5--9, but appear to overestimate the luminosity or underestimate the
effective temperature at L0--5.}
    {We classify six faint Pleiades brown dwarfs as early to mid L-type objects
using low-resolution near-infrared spectra. We compare their properties to
field dwarfs and theoretical models and estimate their masses to be in the range
0.025--0.035~$M_{\odot}$.}

   \maketitle
%

\section{Introduction}

About 600 field L-type dwarfs have been spectroscopically identified to
date.\footnote{http://spider.ipac.caltech.edu/staff/davy/ARCHIVE/index.shtml,\\
2010 May 17.} Most of them have been discovered by large sky-area surveys
such as 2MASS \citep{skrutskie2006} and SDSS \citep{abazajian2009}. Spectroscopy
has allowed the determination of physical properties and the derivation of
effective temperatures (in the range $\sim$2300--1400~K; e.g.\
\citealt{leggett2001}), and parallax studies have provided distances and
luminosities for about 10\% of these objects, indicating that they are most
probably very low mass stars or massive brown dwarfs. For individual field
L-dwarfs, an estimation of mass from the luminosity and effective temperature
requires additional information on age, which is in general difficult to obtain.
This limits our ability to determine the field substellar mass function
and formation history
\citep{reid1999,chabrier2002,chabrier2003,kroupa2003,burgasser2004c,allen2005,pinfield2006}.

\begin{table*}
\caption{Spectroscopic observations.} \label{log}           
\centering          
\begin{tabular}{c c c c c c c c c c}
\hline\hline       
Object      & $J$\tablefootmark{b}	      & Opt.&  Instrument\tablefootmark{d}   & Date	& $T_{\rm exp}$\tablefootmark{e} & Airmass & $\Delta$airmass$_{\rm T-S}$ & S/N\tablefootmark{f} & Near-IR	    \\ 
            &  (mag)	                      & SpT\tablefootmark{c}		    &		                 &	& (min) 	                 &	   &		                 &                      & SpT\\  		      
\hline	    
\object{PPl~1}       & 15.36$\pm$0.01  & M6.5    	&  NICS($JHK$)    & 2002 Oct 31 & 8$\times$2         & 1.08       & 0.14 & 50		 & M7.0$\pm$0.5  \\				   
\object{Calar~3}     & 16.08$\pm$0.01  & M8      	&  NICS($JHK$)    & 2002 Oct 31 & 8$\times$2         & 1.15       & 0.07 & 49		 & M8.0$\pm$0.5  \\				   
\object{Teide~1}     & 16.21$\pm$0.01  & M8      	&  NICS($JHK$)    & 2002 Nov 27 & 8$\times$2         & 1.03       & 0.73 & 32		 & M8.0$\pm$0.5  \\				   
\object{BRB~17}      & 17.41$\pm$0.03  & ...		&  LIRIS($zJ,HK$) & 2006 Dec 28 & (12, 18)$\times$2  & 1.74, 1.41 & 0.44, 0.89 & 4, 6	 & L0.0$\pm$1.0 \\                            
\object{PLIZ~28}     & 17.60$\pm$0.04  & ...		&  NICS($JHK$)    & 2005 Nov 20 & 15$\times$2	     & 1.11       & 0.57 & 18		 & L0.0$\pm$1.0 \\                            
\object{PLIZ~35}     & 18.07$\pm$0.05  & ...		&  NICS($JHK$)    & 2005 Sep 18 & 65$\times$0.6      & 1.03       & $-$0.03 & 15	 & L2.0$\pm$1.0 \\                                
\object{BRB~21}      & 18.14$\pm$0.06  & ...		&  NICS($JHK$)    & 2005 Sep 18 & 80$\times$0.5	     & 1.21       & 0.15 & 7		 & L3.0$\pm$1.0 \\                                
\object{BRB~23}      & 18.23$\pm$0.04  & ...		&  NICS($JHK$)    & 2007 Feb 7  & 10$\times$5	     & 1.33       & $-$0.27 & 14	 & L3.5$\pm$1.0 \\                                
\object{BRB~29}      & 18.69$\pm$0.10  & ...		&  LIRIS($HK$)    & 2006 Dec 29 & 12$\times$2	     & 1.09       & $-$0.02 & 3		 & L4.5$\pm$1.5 \\                                       
\hline	    
J0829\tablefootmark{a} & 12.80$\pm$0.03  & L2V     	&  LIRIS($zJ,HK$) & 2006 Dec 28 &(4, 6)$\times$1.5   & 1.38, 1.36 & 0.80, 0.94 & 29, 37& L2.0$\pm$1.0  \\				       
\hline                  
\end{tabular}
\begin{flushleft} 
\tablefoottext{a}{\object{SSSPM~J0829$-$1309}.}
\tablefoottext{b}{From the Galactic Clusters Survey (GCS) component of UKIDSS
\citep[][sixth data release]{lawrence2007} for the Pleiades members and from
2MASS \citep{skrutskie2006} for the field object.} 
\tablefoottext{c}{From \citet{martin1996} for the three Pleiades objects and from
\citet{scholz2002} and \citet{lodieu2005} for the field object.}
\tablefoottext{d}{A~slit width of 1.0~arcsec was used, except for Calar~3 and
PPl~1 (1.5~arcsec). The NICS/Amici spectra have a constant resolving power
$R=\lambda/\Delta\lambda\approx50$ in the range 0.8--2.5~$\mu$m (dispersion
3--10~nm~pix$^{-1}$); for Calar~3 and PPl~1, $R\approx33$. The LIRIS $zJ$-band
(0.887--1.531~$\mu$m) spectra have $R\approx960$ at $\lambda\approx1.17$~$\mu$m
and a dispersion of 0.61~nm~pix$^{-1}$, whereas the $HK$-band
(1.388--2.419~$\mu$m) spectra have $R\approx945$ at $\lambda\approx1.83$~$\mu$m
and a dispersion of 0.97~nm~pix$^{-1}$.}
\tablefoottext{e}{Dithers $\times$ detector integration time.} 
\tablefoottext{f}{Pseudo-continuum signal-to-noise ratios are measured in the
interval 2.14--2.24~$\mu$m, or for the LIRIS zJ-band data, in the interval
1.28--1.32~$\mu$m.} 
\end{flushleft}
\end{table*}

An empirical determination of the evolution of effective temperature and
luminosity with age for L-type dwarfs could be obtained by identifying such
objects in stellar associations of various ages. Brown dwarf- and planetary-mass
candidates of (mostly early) L-type have been studied photometrically and
spectroscopically in the \object{Serpens} \citep{lodieu2002}, \object{Ophiuchus}
\citep{jayawardhana2006}, \object{Chamaeleon~I}
\citep{luhman2006,luhman2008b,luhman2008c,schmidt2008}, \object{Chamaeleon~II}
\citep{jayawardhana2006b,allers2007}, \object{Taurus}
\citep{itoh2005,luhman2009}, \object{Trapezium}
\citep{lucas2001,lucas2006,weights2009}, and \object{Lupus~1} clouds
\citep{neuhauser2005,mcelwain2007}, the \object{$\sigma$~Orionis} open cluster
\citep{zapateroosorio1999Or,zapateroosorio2000,bejar2001,martin2001,barradoynavascues2001b,barradoynavascues2002or,mcgovern2004},
and the \object{TW~Hydrae} \citep{gizis2002,chauvin2004,looper2007},
\object{Upper~Scorpius} \citep{lodieu2008,lafreniere2008,bejar2008}, and
\object{Tucana--Horologium} associations \citep{chauvin2005b}. All these
star-forming regions are very young, with ages of a few Myr to a few tens of
Myr. Finally, several L-type dwarfs characterized photometrically and
spectroscopically have been associated with moving groups, such as the all-sky,
400$\pm$100~Myr old Ursa Major moving group
\citep{bannister2007,jameson2008a,casewell2008}.

In this context, the \object{Pleiades} open cluster is well suited to provide a
reference sequence for L- and T-dwarfs at an age of $\sim$120 Myr. Proper motion
studies \citep{moraux2003,bihain2006,lodieu2007} have shown that there is a
significant population of low-luminosity members of the cluster that, based on
their optical and near-infrared colours, are likely to be of  spectral type L.
\citet{casewell2007} have extended the brown dwarf search beyond these objects
with some candidate members of  spectral type T. Here, we report near-infrared
spectroscopy for six of the nine L-type candidates confirmed as Pleiades proper
motion members by \citet{bihain2006}, which allows us to study the relationship
between spectral type and luminosity for L dwarfs at a homogeneous age
and solar metallicity.

\section{Observations and data reduction}

Low-resolution spectra of four candidate L-type Pleiades brown dwarfs and three
well-known late M-type Pleiads were obtained  with the Near Infrared Camera 
Spectrometer (NICS) and its Amici prism, mounted on the 3.6~m Telescopio
Nazionale Galileo (TNG; Observatorio del Roque de Los Muchachos --ORM--,
Spain).  Additional spectra of two other candidate L-type Pleiades brown dwarfs
and an L2-type field dwarf were obtained with the Long-Slit Intermediate
Resolution Infrared Spectrograph (LIRIS), mounted at the 4.2~m  William Hershel
Telescope (WHT; ORM). Table~\ref{log} lists the object names, $J$-band
magnitudes, optical spectral types (when available), instruments (and bandpass
coverage), observing-night dates, exposure times, mean airmasses, differences in
mean airmasses between telluric standard stars (T) and the science objects (S),
and pseudo-continuum signal-to-noise ratios. The weather conditions were clear,
except on the nights of 2002 November 27 and 2006 December 28, when some clouds
were present. The seeing was in the range 0.6--1.5~arcsec. Small dithers over
two or three different positions along the slits were performed to allow for sky
subtraction. Spectra of hot, B9--A1V stars were obtained during each observing
night, except in 2002, when the spectra of a DAO white dwarf and a K1 dwarf were
obtained, relatively close in coordinates to the scientific targets, for the
correction of the instrumental response and the telluric absorption.

The raw spectroscopic data were reduced using standard routines within the {\tt
IRAF\footnote{IRAF is distributed by the National Optical Astronomy
Observatories, which are operated by the Association of Universities for
Research in Astronomy, Inc., under cooperative agreement with the National
Science Foundation.}} environment. Amici spectra were sky-subtracted,
flat-fielded, aligned, combined, optimally extracted, and wavelength calibrated
using the look-up table provided in the instrument's
webpage\footnote{http://web.archive.org/web/20071126094610/http://www.tng.iac.es/
\\ instruments/nics/spectroscopy.html} and the deep telluric absorption
features. LIRIS raw images were corrected for pixel mapping and row cross-talk,
and the spectra were sky-subtracted, flat-fielded, wavelength calibrated (using
vacuum wavelengths of Ar arc lamps), shifted, and combined using routines in the
{\tt LIRISDR} package developed by J.~A.~Acosta-Pulido before they were
optimally extracted. The instrumental response and telluric bands were removed
dividing by the ``telluric'' spectra and multiplying by black-body spectra with
the same effective temperature. The intrinsic lines in the LIRIS hot-star
spectra were  removed before dividing  the science spectra.  A vacuum-to-air
correction was applied to the LIRIS spectra, using the default standard
temperature and pressure. As indicated in the column
$\Delta$airmass$_{\rm T-S}$ of Table~\ref{log}, some of our science spectra were
acquired at airmasses quite different to the telluric spectra, thus yielding a
poor correction of telluric bands. The reduced Amici spectrum of BRB~23 was cut
longwards of 2.245~$\mu$m because the level of the  sky counts entered the
non-linear regime of the detector. The LIRIS $zJ$ and $HK$ spectra were flux
calibrated using published $JH$-band photometry, allowing us to combine both
parts.


\section{Spectral types}\label{spt}

\begin{figure}[ht!] \resizebox{\hsize}{!}{\includegraphics{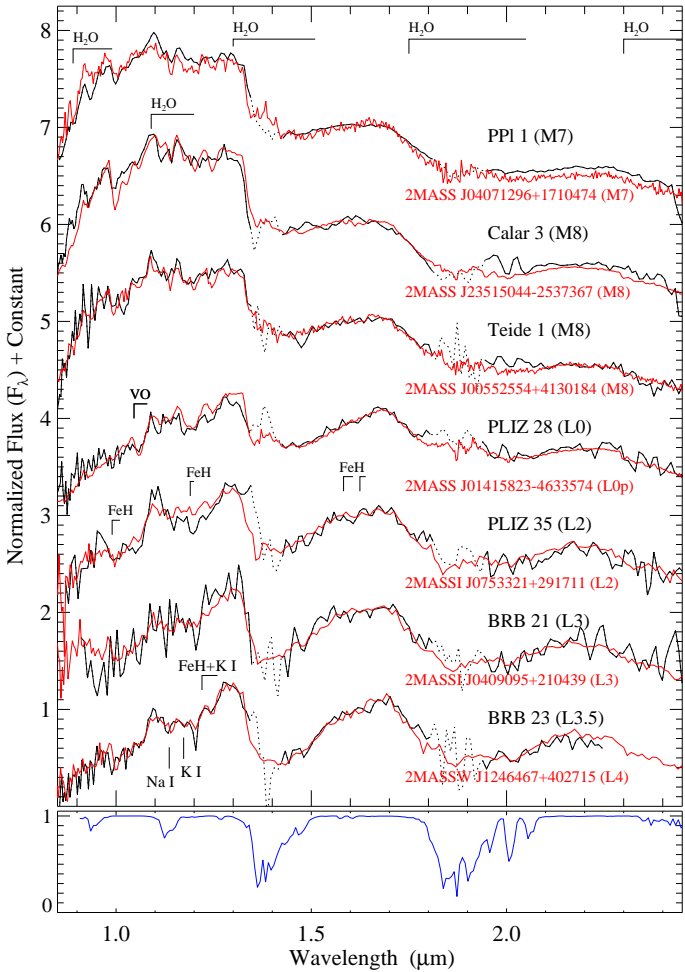}}
\caption{NICS/Amici low-resolution 0.85--2.40~$\mu$m spectra of Pleiades
low-mass members (black lines). The spectra are in air wavelengths, normalized
to the average flux between 1.56 and 1.74~$\mu$m, and offset by constants. The
main regions affected by tellurics are represented by dotted lines. Field dwarf
spectra are overplotted in red and their references are:
\object{2MASS~J04071296$+$1710474} and \object{2MASS~J00552554$+$4130184}
\citep{burgasser2004b}, \object{2MASS~J23515044$-$2537367}
\citep{burgasser2008}, \object{2MASS~J01415823$-$4633574}
\citep{kirkpatrick2006}, \object{2MASSI~J0753321$+$291711},
\object{2MASSI~J0409095$+$210439}, and \object{2MASSW~J1246467$+$402715}
\citep{testi2009}. A model atmospheric transmission rebinned to a dispersion of
7~nm pix$^{-1}$ is represented in blue in the lower panel.} \label{amici}
\end{figure}

\begin{figure}[ht!] \resizebox{\hsize}{!}{\includegraphics{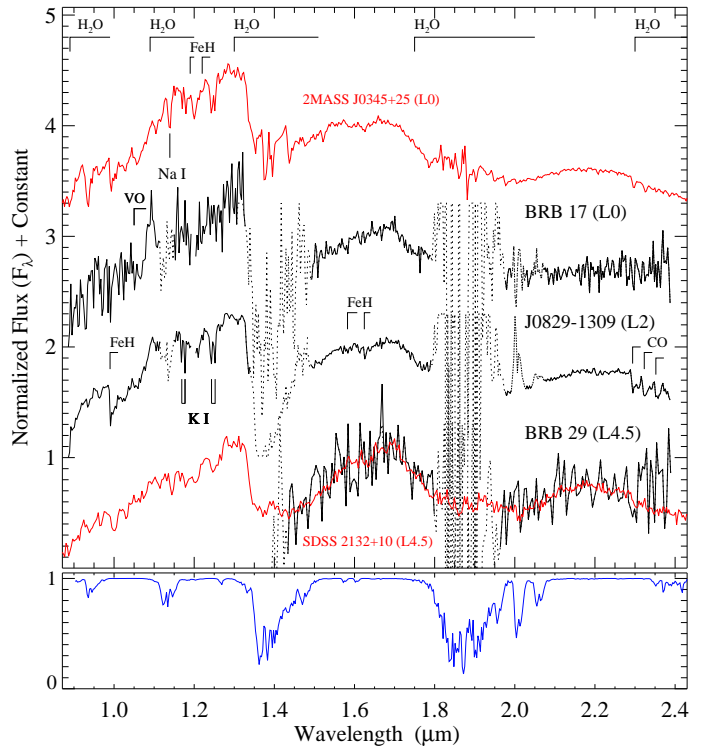}}
\caption{LIRIS intermediate-resolution spectra of the Pleiades low-mass members
BRB~17 and BRB~29 and the field dwarf SSSPM~J0829$-$130 (black lines). The
spectra are in air wavelengths, normalized by the average flux between 1.56 and
1.74~$\mu$m, rebinned to dispersions of 3 (BRB~17 and SSSPM~J0829$-$1309) and
5~nm~pix$^{-1}$ (BRB~29), and offset by constants. The main regions affected by
tellurics are represented by dotted lines. The spectra of the field dwarfs
\object{2MASS~J03454316$+$2540233} (composite spectrum of \citealt{leggett2001},
rebinned to 3~nm pix$^{-1}$) and \object{SDSS~J213240.36$+$102949.4}
\citep[$R\sim150$;][]{chiu2006} are shown in red. A model atmospheric
transmission rebinned to 3~nm pix$^{-1}$ is represented in blue in the lower
panel.} \label{liris} \end{figure}

In Figs~\ref{amici} and \ref{liris}, we show the reduced Amici and LIRIS
spectra, respectively, and also a model atmospheric transmission
\citep{hammersley1998}. The main absorption features of our target spectra are
indicated (see \citealt{cushing2005} for details). As expected for late M- and
L-type objects, the spectra display intense water absorption bands at 1.3--1.51
and 1.75--2.05~$\mu$m. Also, as expected for lower effective
temperatures and dustier atmospheres \citep[see Fig.~2 in][]{tsuji1996a}, the
fainter objects have a flux at 0.85--1.3~$\mu$m that is decreased compared to
that at $H$ or $K$ band.

\begin{figure}[ht!] \resizebox{\hsize}{!}{\includegraphics{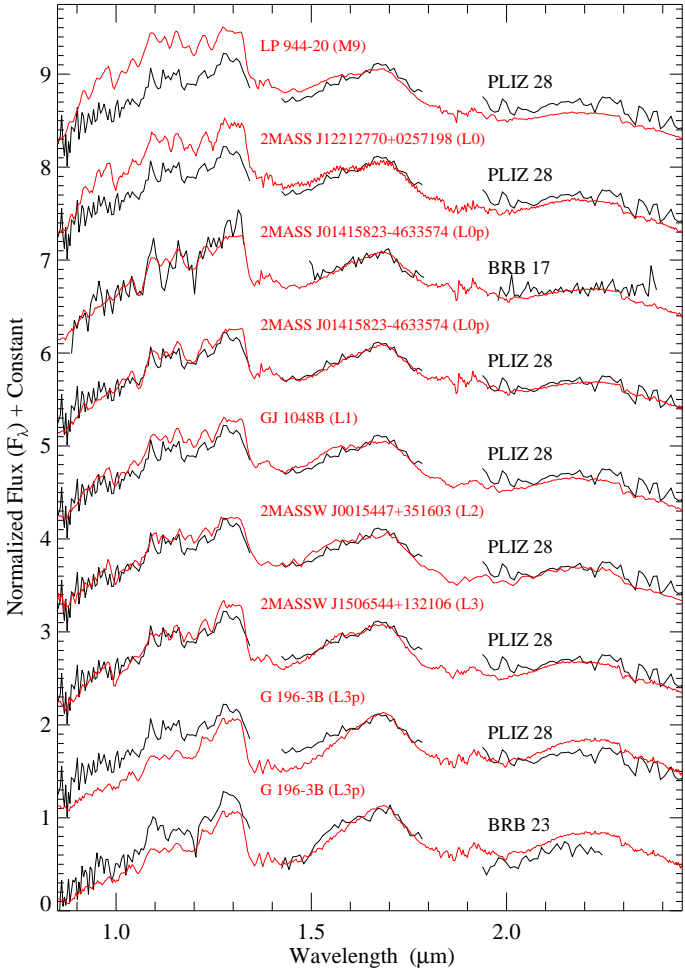}}
\caption{Spectra of Pleiades L-type brown dwarfs compared to field dwarf spectra
(overplotted in red). The spectra are normalized to the average flux between
1.56 and 1.74~$\mu$m and offset by constants.  The LIRIS spectrum of BRB~17 is
rebinned to a dispersion of 7~nm pix$^{-1}$. The references of the field spectra
are: \object{LP~944-20}, \object{2MASS~J12212770$+$0257198}, and
\object{GJ~1048B} \citep{burgasser2008}, \object{2MASS~J01415823$-$4633574}
\citep{kirkpatrick2006}, \object{2MASSW~J0015447$+$351603} \citep{testi2001}, 
\object{2MASSW~J1506544$+$132106} \citep{burgasser2007}, and \object{G~196-3B}
\citep{allers2010}.} \label{comp} \end{figure}

\begin{table*}
\caption{Field-dwarf- and gravity-independent spectral indices\tablefootmark{a} 
(left and right parts of the table, respectively).}         
\label{ind}      
\centering          
\begin{tabular}{c c c c c c c | c c c c}
\hline\hline       
Object   & sHJ  & sKJ  & sH$_{2}$O$^{J}$ & sH$_{2}$O$^{H1}$ & sH$_{2}$O$^{B}$ & $<$SpT$>$\tablefootmark{b} &  H$_{2}$O 1.5~$\mu$m &  H$_{2}$O & H$_{2}$O-2 & $<$SpT$>$\tablefootmark{b}  \\ 
         & {\scriptsize (1.265--1.305,}  & {\scriptsize (1.265--1.305,}  & {\scriptsize (1.09--1.13,}   & {\scriptsize (1.45--1.48,} & {\scriptsize (1.47--1.49,} &  & {\scriptsize (1.46--1.48,} &  {\scriptsize (1.492--1.502,}   & {\scriptsize (2.035--2.045,} &  \\ 
         & {\scriptsize 1.60--1.70)}    & {\scriptsize 2.12--2.16)}    & {\scriptsize 1.265--1.305)} & {\scriptsize 1.60--1.70)} & {\scriptsize 1.59--1.61)} &  & {\scriptsize 1.57--1.59)} &  {\scriptsize 1.55--1.56)} & {\scriptsize 2.145--2.155)} &  \\ 
\hline
BRB~17   & 0.23; L2.4 & 0.63; L2.6 & ...& ... & ... &  (L2.5)   & ... & 1.22; L0.6 & 0.95; M8.4  & (M9.5) \\
PLIZ~28  & 0.12; L4.5 & 0.51; L4.0 & 0.21; L3.0  & 0.35; L2.0 & 0.78; L0.8  & (L3)   & 1.30; L1 & 1.17; M9.5 & 0.88; L0.3  & (L0.5) \\
PLIZ~35  & 0.21; L2.7 & 0.66; L2.2 & 0.09; L1.2  & 0.37; L2.3 & 0.79; L0.6  & (L2)   & 1.31; L1 & 1.36; L4.0 & 0.80; L2.6  & (L2.5) \\
BRB~21   & 0.19; L3.1 & 0.55; L3.5 & 0.35; L5.2  & 0.38; L2.4 & 0.67; L3.6  & (L3.5) & 1.39; L2 & 1.42; L5.2 & 0.89; L0.0  & (L2.5) \\
BRB~23   & 0.18; L3.4 & 0.55; L3.5 & 0.29; L4.3  & 0.58; L5.0 & 0.58; L6.2  & (L4.5) & 1.67; L7 & 1.51; L7.2 & 0.79; L2.8  & (L5.5) \\
BRB~29   & ...        & ...        & ...         & 0.65; L5.8 & 0.68; L3.4  & (L4.5) & 1.56; L5 & 1.25; L1.3 & 0.68; L5.6  & (L4) \\
\hline                  
\end{tabular}
\begin{flushleft}
\tablefoottext{a}{Spectral ranges in $\mu$m are indicated between parentheses.}
\tablefoottext{b}{Average spectral types derived from the spectral indices
are approximated to the nearest half subclasses.}
\end{flushleft}
\end{table*}

The Amici spectra were classified spectroscopically by comparison with 
spectra of the Amici \citep{testi2001,testi2009} and SpeX Prism
libraries\footnote{http://www.browndwarfs.org/spexprism} (Adam Burgasser), with
resolving power $R\sim100$ and $JHK$-bandpass coverage. The LIRIS spectra were
classified similarly using in addition the $R$~$\sim$~150--600 spectra from the
L- and T-type dwarf data archive of Sandy
Leggett.\footnote{http://staff.gemini.edu/$\sim$sleggett/LTdata.html} The first,
second, and third libraries adopt L spectral types from the far-red optical
(classification scheme of \citealt{kirkpatrick1999b}), the far-red optical or
near-infrared, and the near-infrared (classification scheme of
\citealt{geballe2002}, which agrees in the L0--L5 range with the optical schemes
of \citealt{martin1999} and \citealt{kirkpatrick1999b}), respectively. For the
objects of the second and third libraries, we preferentially adopted optical
spectral types, when available\footnote{http://www.dwarfarchives.org}.

We classified the spectra first visually, considering the change in the 
continuum shape from the earlier to the later spectral types (e.g., the
0.85--1.3~$\mu$m region and the H$_{2}$O absorption bands), and then by
chi-squared minimization  for optimization. The errors were obtained by
accounting for the plausible earliest- and latest-type field dwarf spectra
matching those of our targets. For BRB~17 and PLIZ~28, however, the spectra
differ notably from those of normal field dwarfs. Figure~\ref{comp} illustrates
this with a sequence of field dwarfs, whose optical-typing references are:
\object{LP~944-20} \citep{kirkpatrick1999b}, \object{2MASS~J12212770$+$0257198}
\citep{reid2008}, \object{2MASS~J01415823$-$4633574}
\citep{kirkpatrick2006,kirkpatrick2008,cruz2009}, \object{GJ~1048B}
\citep{gizis2001}, \object{2MASSW~J0015447$+$351603} \citep{kirkpatrick2000}, 
\object{2MASSW~J1506544$+$132106} \citep{gizis2000}, and \object{G~196-3B}
\citep{cruz2009}. The spectra of BRB~17 and PLIZ~28 match at best the spectrum
of the low surface-gravity field dwarf \object{2MASS~J01415823$-$4633574}
\citep{kirkpatrick2006}. Therefore we adopted its L0$\pm$1 (peculiar) spectral
type. We compared also with G~196-3B, another low surface-gravity field dwarf
\citep{rebolo1998,martin1999,kirkpatrick2001,allers2007,kirkpatrick2008}, but
the match is less good, as with any of the other Pleiades spectra. The latter
compare at best with the spectra of the normal field dwarfs overplotted in
Figs~\ref{amici} and~\ref{liris}. We verified whether the $J-H$ and $H-K$
colours of our Pleiades targets agree with those of the specific field dwarfs
providing the best spectral match. For the former objects we used
UKIDSS\footnote{UKIDSS uses the UKIRT Wide Field Camera (WFCAM;
\citealt{casali2007}) and a photometric system described in \citet{hewett2006}.
The pipeline processing and science archive are described in
\citet{hambly2008}.} photometry and for the latter dwarfs we used 2MASS
photometry converted to the WFCAM photometric system using the transformations
of \citet{hodgkin2009}, or else when available UKIDSS- (2MASSI~J0409095+210439)
or MKO photometry (SDSS~J213240.36$+$102949.4; \citealt{chiu2006}). On average,
the colours agree within the error bars, and, in all cases but two, better than
about 1.5 times the sum of the error bars. 2MASS~J00552554$+$4130184, with
$H-K\approx0.0$~mag, is 0.5~mag bluer than Teide~1, which is in disagreement
with the very similar fluxes in the spectra of the two objects
(Fig.~\ref{amici}). 2MASS~J01415823$-$4633574 is 0.20~mag redder in $H-K$ than
BRB~17.

Our derived spectral types (Table~\ref{log}) indicate that the low-mass brown
dwarfs confirmed by proper motion in \citet{bihain2006} are indeed ultra-cool
dwarfs of spectral type~L. They indicate also that our near-infrared
classification of the late M-type Amici spectra is consistent with
their optical spectral classification. \citet{steele1995b},
\citet{williams1996} and \citet{greissl2007} have already presented
near-infrared low-resolution spectra of Pleiades early M- to L0-type dwarfs,
including \object{PPl~1}, \object{Teide~1}, \object{Roque~33}, and
\object{Roque~25}, which at least qualitatively support the optical
classification.

For the Pleiades L-type spectra, we measured five field-dwarf spectral indices,
in regions less affected by telluric absorption bands: sHJ, sKJ,
sH$_{2}$O$^{J}$, and sH$_{2}$O$^{H1}$ from \citet{testi2001} and sH$_{2}$O$^{B}$
from \citet{reid2001b} . For BRB~17, only the sHJ and sKJ indices could be
measured. We note that the indices depend on narrow regions that are
significantly affected by noise in our data, implying that they are less
reliable than spectral comparisons using the whole regions available. The
average spectral types derived from the field dwarf indices are shown in the
left part of Table~\ref{ind} and agree with our adopted spectral types
(Table~\ref{log}) except for BRB~17 and PLIZ~28. We also measured the H$_{2}$0
1.5~$\mu$m \citep{geballe2002}, H$_{2}$O \citep{allers2007}, and H$_{2}$O-2
\citep{slesnick2004} spectral indices, expected to be gravity-independent in
early to mid L-types. The two latter indices are more appropriate for spectra of
higher spectral resolution than the Amici data. Nevertheless, we find good
agreement between the average spectral types derived from the field-dwarf and
``gravity-independent'' indices. In the case of PLIZ~28 and BRB~17, the spectral
indices independent of the $J$ band -- especially the gravity-independent
indices -- give earlier spectral types, more consistent with the spectral type
L0. This results from the apparently shallower water bands at $\sim$1.4 and
$\sim$1.9~$\mu$m for the relatively depressed continuum at 0.85--1.3~$\mu$m, as
compared to normal field dwarfs. Spectra of improved signal-to-noise ratio and
flux correction will permit us to confirm or refute this unusual
combination of spectral features.

\section{Discussion}

\subsection{Spectroscopic properties}\label{specf}

In Fig.~\ref{liris}, we compare the spectrum of the L0-type BRB~17 with those of
two field dwarfs, the optical L0 standard \object{2MASS~J03454316$+$2540233} of
\citet{kirkpatrick1999b} and the L2-type \object{SSSPM~J0829$-$1309}. In the
blue part, the Pleiades spectrum displays the VO band at $\sim$1.05~$\mu$m,
which appears more clearly than other local features (we note that the FeH band
at 1.2~$\mu$m is not covered because of bad pixel columns in the LIRIS frames).
In the central part ($H$ band), the shape of the continuum  is triangular,
instead of approximately flat-topped as in the comparison spectra. These
peculiarities can also be seen in the lower resolution spectrum of PLIZ~28
(Figs.~\ref{amici}~and~\ref{comp}). Deep VO bands and triangular $H$-band
continuum shapes, as well as weak FeH and CO bands and weak \ion{Na}{i} and
\ion{K}{i} lines, have been observed in spectra of L-type objects in 
star-forming regions and have been associated with lower surface gravity
\citep{martin1998b,lucas2001,mcgovern2004,allers2007}. The triangular $H$-band
continuum shape could be explained by a reduction in the H$_{2}$
collision-induced and H$_{2}$O absorptions in lower surface-gravity and dustier
atmospheres \citep{borysow1997,kirkpatrick2006,mohanty2007}. Possibly, the
entire apparently peculiar continuum of PLIZ~28 and BRB~17 (see Sect.~\ref{spt})
could be explained by more dusty atmospheres \citep[see Fig.~2
in][]{tsuji1996a}.


\subsection{Spectrophotometric properties}\label{spp}

\begin{figure*}[ht!] \resizebox{\hsize}{!}{\includegraphics{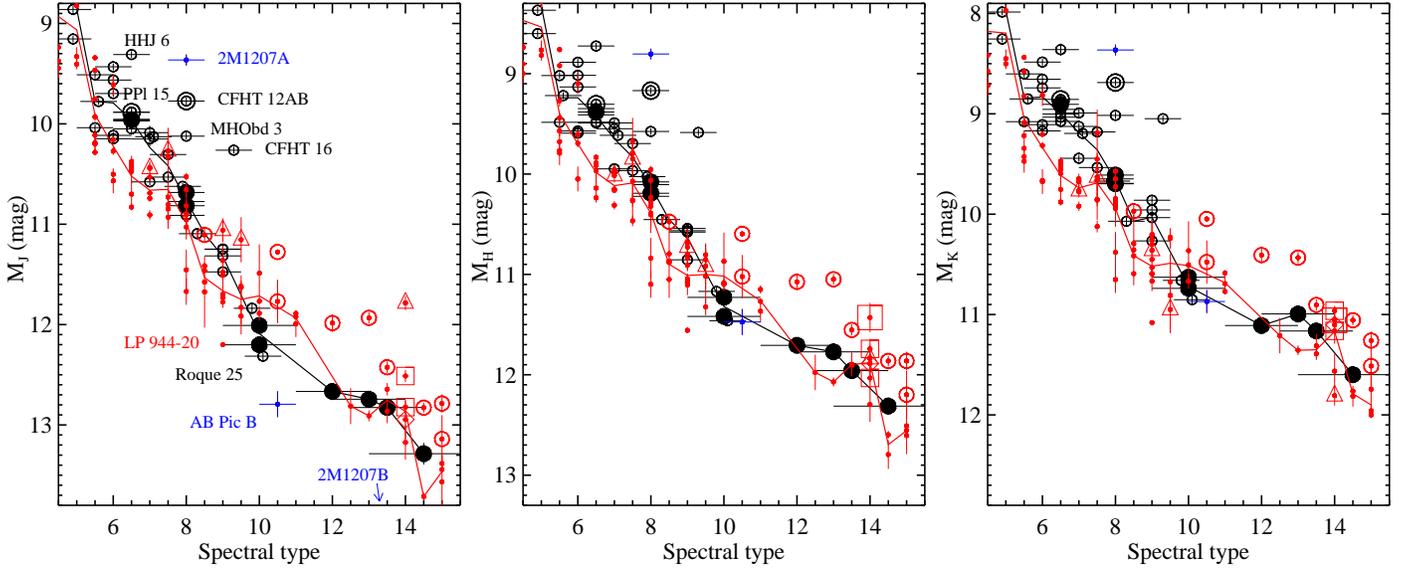}}
\caption{Absolute magnitude $M_J$, $M_H$, and $M_K$ versus spectral type (M6=6,
L0=10) diagrams ({\it left-}, {\it middle-}, and {\it right panels}, respectively), with
Pleiades members ({\it black filled circles} for new data and {\it black small
open circles} for previous data) and field dwarfs ({\it red dots}). Resolved
close binaries are {\it encircled}. Average Pleiades and field values are
represented by the {\it black-} and {\it red solid lines}, respectively. For
illustrative purpose, we highlight four relatively bright field L4-type dwarfs 
(\object{HD~130948B} and C by {\it small squares}, \object{HD~49197B} by {\it
large square}, and \object{2MASSW~J1841086+311727} by {\it diamond}) and we add
five subdwarfs ({\it triangles}) and three low-mass substellar members of very
young associations ({\it blue dots}; we assume an L5 type for 2M1207B). All
three diagrams span 5.0~mag. The photometry is in the MKO-WFCAM system.}
\label{diag} \end{figure*}

We compare the spectrophotometric properties of Pleiades L-type brown dwarfs
mainly with those of field dwarfs of known parallax. We use $I_{\rm C}$- and
UKIDSS $ZYJHK$-band photometry. To our nine Pleiades targets, we add 36 other
low-mass Pleiades stars and brown dwarfs with spectral type measurements
(Table~\ref{sample}). They are cluster members confirmed by proper motion or the
lithium test \citep{rebolo1992}, and many present spectral features of low
surface gravity (see spectroscopic references in the table). We include the
M9-type \object{Roque~4} and the L0-type \object{Roque~25}, which have spectral
features consistent with cluster membership
\citep{zapateroosorio1997b,martin1998b,martin2000,kirkpatrick2008}; Roque~25 has
furthermore a proper motion consistent with cluster membership (see
Appendix~\ref{pm25}). To obtain $JHK$-band absolute magnitudes, we use the
Pleiades revised trigonometric parallax distance of 120.2$\pm$1.9~pc
\citep{vanleeuwen2009,vanleeuwen2009b}. We do not account for the distance
uncertainty due to cluster depth \citep[$\approx$$\pm$0.2~mag towards the
cluster centre,][]{pinfield1998}. We compute average $JHK$-band absolute
magnitudes and $I_{\rm C}-J$, $Z-J$, and $Y-J$ colours in the spectral type
intervals [4.75+i/2,5.25+i/2[ with i=0...19 (where 5.0 corresponds to M5.0, 14.5
to L4.5) and excluding known binaries. We also obtain chi-squared linear fits to
the $J-H$, $H-K$, and $J-K$ colours as a function of spectral type, including
these binaries.

For field dwarfs compiled with trigonometric parallaxes and -- preferentially
optical -- spectral types
\citep{vanaltena1995,dahn2000,dahn2002,henry2004,faherty2009}, we use $I_{\rm
C}$- \citep{leggett2000,dahn2000,dahn2002,henry2004,phan-bao2008} and 2MASS or
MKO $JHK$-band photometry. We convert the 2MASS photometry to the (MKO-)WFCAM
system using the transformations of \citet{hodgkin2009}. For some field dwarfs,
we use available WFCAM $ZYJHK$-band photometry (UKIDSS DDR6) or WFCAM $Z-J$ and
$Y-J$ synthetic colours \citep{hewett2006,rayner2009}. We cautiously assume
errors of 0.10~mag and 0.05~mag in the colours of the former and latter studies,
respectively. Late M- and L-type close binaries resolved by imaging \citep[see
references in][]{faherty2009} are accounted for in the comparison. We compute
average absolute magnitudes, average colours and linear fits in a similar way to
the Pleiades sample, but using values with total errors smaller than 0.4~mag.
For illustrative purpose, we highlight four relatively bright L4-type field
dwarfs (\object{HD~130948B} and C, \citealt{potter2002,dupuy2009};
\object{HD~49197B}, \citealt{metchev2004}; \object{2MASSW~J1841086+311727},
\citealt{kirkpatrick2000}) and we add five subdwarfs associated with the
thick-disc or halo population \citep{bowler2009}, three low-mass substellar
members of relatively young associations (\object{2M1207AB},
\citealt{chauvin2004,mohanty2007,ducourant2008}; \object{AB~Pic~B},
\citealt{chauvin2005b,bonnefoy2010}), and three field dwarfs without parallaxe
measurements (\object{2MASS~J01415823$-$4633574}, G\,196--3B and the candidate
young L4.5-type \object{2MASS~J18212815+1414010}, \citealt{looper2008}).

\begin{figure*}[ht!] \resizebox{\hsize}{!}{\includegraphics{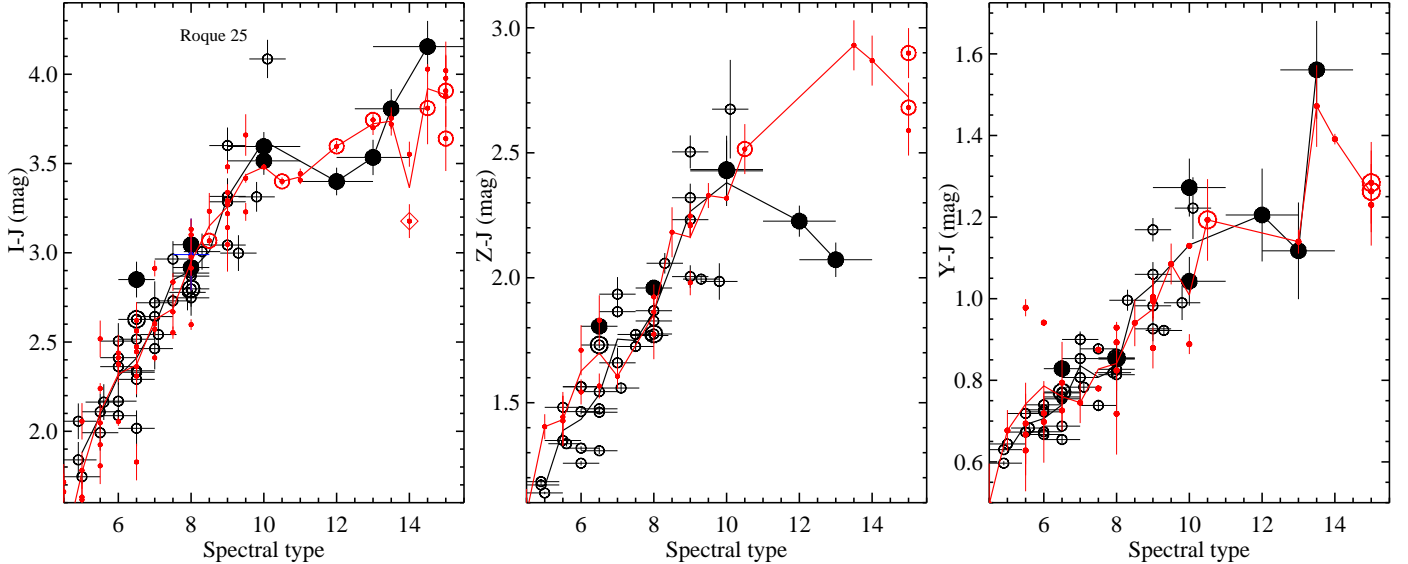}}
\caption{$I_{\rm C}-J$, $Z-J$, and $Y-J$ versus spectral type diagrams
({\it left-}, {\it middle-}, and {\it right panels}, respectively). Same symbol
definitions as in Figure~\ref{diag}.} \label{diagIZYJ} \end{figure*}

\begin{figure*}[ht!] \resizebox{\hsize}{!}{\includegraphics{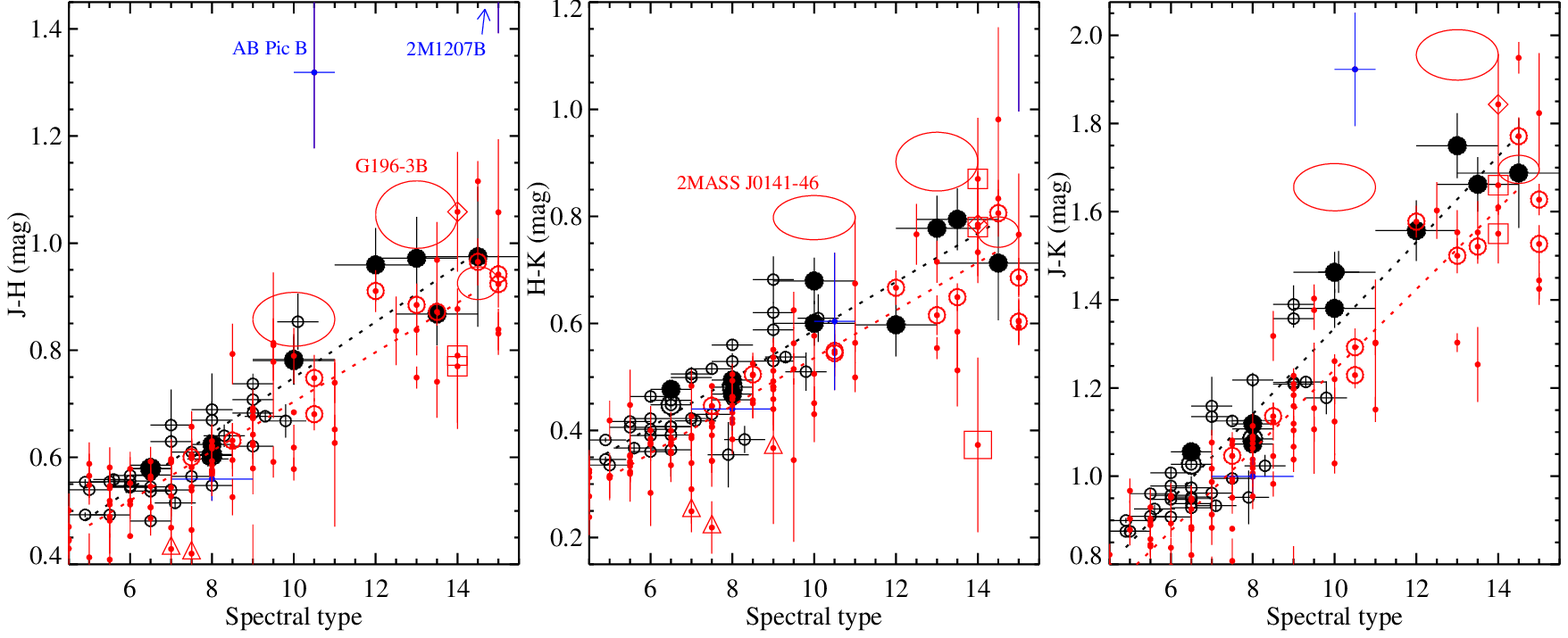}}
\caption{$J-H$, $H-K$, and $J-K$ versus spectral type diagrams ({\it left-},
{\it middle-}, and {\it right panels}, respectively). Same symbol definitions as
in Fig.~\ref{diag}. Pleiades and field chi-squared linear fits to the colours as
a function of spectral type are represented by the {\it black-} and {\it red
dashed lines}, respectively. For illustrative purpose, we add the low surface
gravity field dwarfs \object{2MASS~J01415823$-$4633574} and G\,196--3B and the
candidate young field dwarf \object{2MASS~J18212815+1414010}, represented by
error {\it ellipses} for clarity.} \label{diagJHK} \end{figure*}

As claimed in \citet{bihain2006}, the Pleiades and field L-type dwarfs may have
similar spectral energy distributions and luminosities, and thus possibly
similar (but not necessarily equal) radii, based on the apparent overlapping of
these objects in the $J$ versus $I-J$, $J-H$ and $J-K$ colour--magnitude
diagrams. Indeed, our near-infrared absolute magnitude versus spectral type
diagrams (Fig.~\ref{diag}) indicate that unlike the Pleiades M-type dwarfs, the
L-type dwarfs appear no brighter than a few tenths of magnitude than their field
counterparts, especially in the $J$~band. Nevertheless, if we adopt as in
\citet{bihain2006} the larger Pleiades distance of 133.8$\pm$3.0~pc
\citep{percival2005} -- obtained by main sequence fitting in the optical and
near-infrared using field dwarfs -- the Pleiades L-type dwarfs would be brighter
by 0.23~mag, implying less similar luminosities to those of the field dwarfs.
Besides, the diagrams of Figs.~\ref{diagIZYJ}~and~\ref{diagJHK} show no
significant differences between the colours of Pleiades and field L-type
dwarfs.

At the M/L transition (see Fig.~\ref{diag}), we note that four Pleiades
brown dwarfs (\object{NPL~40}, \object{BRB~17}, \object{PLIZ~28} and
\object{Roque~25}) appear to have a relatively lower flux towards the $J$ band,
which could be due to a too early spectral typing or some atmospheric or
interstellar extinction. The effect is stronger for the $\sim$30~Myr old
low-mass brown dwarf or massive planet \object{AB~Pic~B}, and even stronger for
the later-type $\sim$8~Myr few-Jupiter-mass
\object{2M1207B}\footnote{\object{2M1207B} is less luminous than the Pleiades
and field dwarf sequences, by about 3, 2, and 1 magnitudes in the $J$, $H$, and
$K$~bands, respectively, assuming a spectral type L5.}, for which
ongoing accretion/formation processes have been proposed
\citep{mohanty2007,mamajek2007,ducourant2008,patience2010}. Finally the very
nearby (5~pc) M9-type field brown dwarf \object{LP~944-20}, with lithium
absorption and an age estimated to be of several hundred Myr
\citep{tinney1998,ribas2003} appears also to be fainter, but equally in the $HK$
bands as in the $J$ band, an effect that could be real or related to
uncertainties in the parallax measurement.

The near-infrared colour diagrams of Fig.~\ref{diagJHK} indicate that late M-
and L-type Pleiades dwarfs are marginally redder in $J-H$ and $H-K$ than their
field counterparts. The difference reaches 0.11$\pm$0.20~mag in $J-K$ and
at spectral type L3, using the chi-squared linear fits. If we consider only the
field dwarfs with UKIDSS photometry (10 M6--L4-type dwarfs), the $J-K$
difference at L3 increases by 0.08~mag. This suggests that the slightly redder
colour is not due to some bias in the photometric transformations of
\citet{hodgkin2009}. It could be associated with both:

\begin{enumerate}

\item The reddening in the line of sight of the faint Pleiades brown dwarfs. The
$E(B-V)$ values could be in the range 0.02--0.14~mag \citep{taylor2008}, which
using the $E(J-K_{\rm s})/E(B-V)$ relation of \citet{an2007}, would lead to
small infrared excesses $E(J-K_{\rm s})=0.01-0.09$~mag for an intrinsic colour
$(J-K_{\rm s})_0=1.5$~mag.

\item The youth of the Pleiades brown dwarfs. Indeed, the $J-K$ colour is found
to be larger, in average, for L-type field dwarfs with smaller velocity
dispersion, and thus of younger age \citep{faherty2009,schmidt2010}. The 0.1~Gyr
old Pleiads would likely have lower surface gravities than those of the older
field dwarfs, which furthermore could have lower metallicities than the Pleiades
([Fe/H]$_{\rm Pleiades}$~=~0.03$\pm$0.05~dex,
\citealt{funayama2009,soderblom2009}). Both lower surface gravities and higher
metallicities may cause lower H$_{2}$ collision induced absorption or dustier
atmospheres, and therefore larger $J-K$ colours
\citep{linsky1969,saumon1994,borysow1997,mohanty2007}. Also, several field
L-type dwarfs with spectral features of low surface gravity have very red
near-infrared colours (see Fig.~\ref{diagJHK} and
\citealt{kirkpatrick2008,cruz2009}).

\end{enumerate}


\subsection{Hertzsprung--Russell diagram} 

\begin{figure}[ht!] \resizebox{\hsize}{!}{\includegraphics{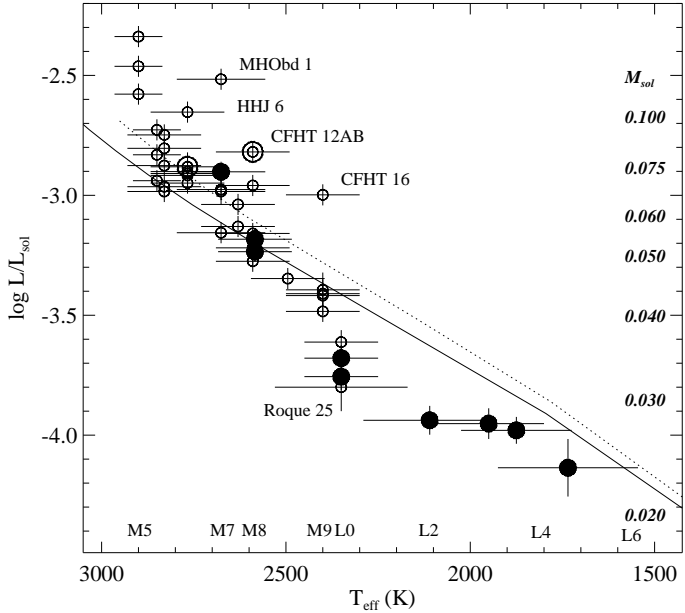}}
\caption{Hertzsprung--Russell diagram of Pleiades low-mass stars and brown
dwarfs. Same symbol definitions as in Figure~\ref{diag}. Theoretical 120~Myr
isochrones of \citet[][DUSTY]{chabrier2000b} and \citet[][]{burrows1997} are
represented by dotted and solid lines, respectively. Spectral types and
DUSTY theoretical masses are indicated.} \label{hr} \end{figure}

Altogether, the L-type Pleiades and field dwarf sequences appear to be rather
close in the photometric -- spectral-type diagrams.  We build now the
Hertzsprung--Russell (HR) diagram of our sample of 45 Pleiades members. We
convert observed $J$-band magnitudes into luminosities using the open cluster
revised parallactic distance and an average of field dwarf $J$-band bolometric
corrections $BC_J$ \citep[][]{dahn2002,vrba2004} as a function of spectral type.
The $J$ band is chosen because the spectral energy distributions of the Pleiades
objects peak in this band and the $J-K$ colour excesses with respect to their
field dwarf spectral type counterparts are only 0.1--0.2~mag. For the
effective temperatures ($T_{\rm eff}$), we use an average of field dwarf $T_{\rm
eff}$ determinations \citep{basri2000b,dahn2002,vrba2004} also as a function of
spectral type. We provide the luminosity and $T_{\rm eff}$ values for our nine
Pleiades targets in Table~\ref{deriv}, where the error bars account for the
uncertainties in the photometry, cluster distance, spectral types, and the
spread in the field dwarf $BC_J$ and $T_{\rm eff}$. Figure~\ref{hr} shows the
Hertzsprung--Russell diagram. The Pleiades substellar sequence clearly extends
to mid-L spectral types or $T_{\rm eff} \sim 1700$~K. The ``single-object''
sequence has a luminosity dispersion of about 0.1~dex (except at $\sim$M9--L0)
and no apparent temperature gap between 2900 and 1700~K. The sequence of
suspected equal-mass binaries is clearly seen at the warmest temperatures,
whereas below 2400~K there is no obvious hint of such objects, despite the fact
that photometric surveys tend to be biased towards their detection.

We overplot the theoretical 120-Myr isochrones of \citet[][DUSTY
model]{chabrier2000b} and \citet{burrows1997} in Fig.~\ref{hr}. Both models make
similar predictions for luminosity, $T_{\rm eff}$ and mass, and both appear to
reproduce well the slope of the mid- to late-M single-object sequence of the
Pleiades within about 1-$\sigma$ uncertainty. There is, however, a discrepancy
between theory and observations in the L-type regime ($\le$2400~K): our objects
appear 100--400~K warmer (1--3 spectral subclasses earlier) or 0.1--0.4~dex less
luminous than the isochrones. Using the $K$ band and the field dwarf bolometric
correction from \citet{golimowski2004} increase by $\la$0.10~dex the
luminosities of the Pleiades L-type objects with the largest $J-K$ colour
excess, thus only slightly reducing the discrepancy. It is uncertain whether the
field dwarf bolometric correction- and $T_{\rm eff}$--spectral type relations
may be valid for the Pleiades L-type brown dwarfs because of gravity dependence.
However, the latter appear to have spectral energy distributions and
luminosities that are in general very close to those of their field dwarf
spectral type counterparts. Thus, the discrepancy in the Hertzsprung--Russell
diagram may also be due to incorrect model predictions, such as an
underestimation of the contraction rate of the Pleiades L-type dwarfs.

\begin{table}
\caption{Luminosities, effective temperatures, and masses for our Pleiades targets.}          
\label{deriv}      
\centering          
\begin{tabular}{c c c c c}
\hline\hline       
Object            & SpT           & $\log{L/L_{\odot}}$   &  $T_{\rm eff}$  & $M$ \\ 
                  &               & (dex)                 & (K)             & ($M_{\odot}$) \\                 
\hline	    
\object{PPl~1}    & M6.5$\pm$0.5  & $-$2.90$\pm$0.04 & 2676$\pm$120 & $\sim$0.074 \\                                
\object{Calar~3}  & M8.0$\pm$0.5  & $-$3.18$\pm$0.04 & 2584$\pm$100 & $\sim$0.054 \\                                
\object{Teide~1}  & M8.0$\pm$0.5  & $-$3.24$\pm$0.04 & 2584$\pm$100 & $\sim$0.052 \\                                
\object{BRB~17}   & L0.0$\pm$1.0  & $-$3.68$\pm$0.06 & 2350$\pm$100 & $\sim$0.035 \\                            
\object{PLIZ~28}  & L0.0$\pm$1.0  & $-$3.76$\pm$0.06 & 2350$\pm$100 & $\sim$0.033 \\                            
\object{PLIZ~35}  & L2.0$\pm$1.0  & $-$3.94$\pm$0.06 & 2110$\pm$180 & $\sim$0.028 \\                                
\object{BRB~21}   & L3.0$\pm$1.0  & $-$3.95$\pm$0.06 & 1950$\pm$150 & $\sim$0.028 \\                                
\object{BRB~23}   & L3.5$\pm$1.0  & $-$3.98$\pm$0.06 & 1875$\pm$150 & $\sim$0.027 \\                                
\object{BRB~29}   & L4.5$\pm$1.5  & $-$4.14$\pm$0.12 & 1735$\pm$190 & $\sim$0.024 \\                                       
\hline                  
\end{tabular}
\end{table}

The luminosity range of the spectroscopic sample as compared to the theoretical
luminosities of the 120~Myr DUSTY isochrone (Fig.~\ref{hr}) would imply a mass
range of $\sim$0.025--0.1~$M_{\odot}$ (without accounting for the brightest,
probable binaries). In Table~\ref{deriv}, we list individual masses for our
targets, obtained by interpolating linearly between the model points using our
estimated luminosities. The faintest and coolest brown dwarf is \object{BRB~29},
with an estimated mass of 25\,$\pm$\,8~$M_{\rm Jup}$.

\section{Conclusions}

We obtained near-infrared NICS/Amici and LIRIS spectra of six $J=17.4$--18.8~mag
candidate L-type brown dwarfs of the $\sim$120~Myr old Pleiades cluster, as well
as Amici spectra of three other members that are well-known late M-type brown
dwarfs. We confirm that the former objects have early- to mid-L spectral types
and also confirm the optical spectral types reported for the latter objects.
\object{PLIZ~28} and \object{BRB~17}, the two earliest L-type brown dwarfs,
present a triangular $H$-band continuum shape and a noticeable VO band,
associable to low surface gravity, dust and youth. The Pleiades L-type
brown dwarfs appear to have similar absolute magnitudes and colours as their
field counterparts, but have slightly redder near-infrared colours, possibly
related to both reddening and youth.


We build the Hertzsprung--Russell diagram of our Pleiades sample using field
dwarf relations and find good agreement with theoretical models for the
M5.5--9~type objects. However, for the L0--5~type brown dwarfs, models  appear
to overpredict the luminosity ($\Delta \log{L/L_{\odot}}=0.1-0.4$~dex) or
underestimate the effective temperature ($\Delta T_{\rm eff}=100$--400~K). It is
also possible that  the adopted field dwarf relations are not truly valid for
$\sim$120~Myr-old objects. Additional optical-to-infrared spectroscopy and
higher resolution imaging of the Pleiades L-type brown dwarfs may allow
us to account more for dust opacities and binarity
\citep[e.g.,][]{knapp2004,burgasser2008}, and refine our estimates of
spectral types, effective temperatures, luminosities, and masses. Current models
predict masses in the range 0.025--0.035~$M_{\odot}$ for the Pleiades L0--5~type
brown dwarfs.

\begin{acknowledgements}


We thank the anonymous referee for her/his valuable comments and suggestions,
and Terry Mahoney for correcting the English text. We thank Katelyn N. Allers
for providing the spectrum of G~196-3B. The study presented here is based on
observations made with the Italian Telescopio Nazionale Galileo (TNG) and the
William Herschel Telescope (WHT) operated on the island of La Palma by the
Fundaci{\'o}n Galileo Galilei of the INAF (Istituto Nazionale di Astrofisica)
and the Isaac Newton Group, respectively, in the Spanish Observatorio del Roque
de los Muchachos of the Instituto de Astrof\'{\i}sica de Canarias. We thank the
TNG for allocation of director's discretionary time to this programme. Based on
observations collected at the German-Spanish Astronomical Center, Calar Alto,
jointly operated by the Max-Planck-Institut f{\"u}r Astronomie Heidelberg and
the Instituto de Astrof\'{\i}sica de Andaluc\'{\i}a (CSIC). We thank Calar Alto
Observatory for allocation of director's discretionary time to this programme.
This research has been supported by the Spanish Ministry of Science and
Innovation (MICINN). This research has benefitted from the SpeX Prism Spectral
Libraries maintained by Adam Burgasser, the L and T dwarf data archive
maintained by Sandy K. Leggett, and the M, L, and T dwarf compendium housed at
DwarfArchives.org and maintained by Chris Gelino, Davy Kirkpatrick, and Adam
Burgasser. This research has made use of SAOImage DS9, developed by Smithsonian
Astrophysical Observatory. This research has made use of the SIMBAD database,
operated at CDS, Strasbourg, France. This publication makes use of data products
from the Two Micron All Sky Survey, which is a joint project of the University
of Massachusetts and the Infrared Processing and Analysis centre/California
Institute of Technology, funded by the National Aeronautics and Space
Administration and the National Science Foundation. This research has made use
of NASA's Astrophysics Data System Bibliographic Services.

\end{acknowledgements}

\bibliographystyle{aa}
\bibliography{/home/gbihain/home_talla/substellar_phd_thesis/astro_ref/astronomy_ref}

\begin{thebibliography}{143}
\expandafter\ifx\csname natexlab\endcsname\relax\def\natexlab#1{#1}\fi

\bibitem[{{Abazajian} {et~al.}(2009){Abazajian}, {Adelman-McCarthy},
  {Ag{\"u}eros}, {Allam}, {Allende Prieto}, {An}, {Anderson}, {Anderson},
  {Annis}, {Bahcall}, {Bailer-Jones}, {Barentine}, {Bassett}, {Becker},
  {Beers}, {Bell}, {Belokurov}, {Berlind}, {Berman}, {Bernardi}, {Bickerton},
  {Bizyaev}, {Blakeslee}, {Blanton}, {Bochanski}, {Boroski}, {Brewington},
  {Brinchmann}, {Brinkmann}, {Brunner}, {Budav{\'a}ri}, {Carey}, {Carliles},
  {Carr}, {Castander}, {Cinabro}, {Connolly}, {Csabai}, {Cunha}, {Czarapata},
  {Davenport}, {de Haas}, {Dilday}, {Doi}, {Eisenstein}, {Evans}, {Evans},
  {Fan}, {Friedman}, {Frieman}, {Fukugita}, {G{\"a}nsicke}, {Gates},
  {Gillespie}, {Gilmore}, {Gonzalez}, {Gonzalez}, {Grebel}, {Gunn},
  {Gy{\"o}ry}, {Hall}, {Harding}, {Harris}, {Harvanek}, {Hawley}, {Hayes},
  {Heckman}, {Hendry}, {Hennessy}, {Hindsley}, {Hoblitt}, {Hogan}, {Hogg},
  {Holtzman}, {Hyde}, {Ichikawa}, {Ichikawa}, {Im}, {Ivezi{\'c}}, {Jester},
  {Jiang}, {Johnson}, {Jorgensen}, {Juri{\'c}}, {Kent}, {Kessler}, {Kleinman},
  {Knapp}, {Konishi}, {Kron}, {Krzesinski}, {Kuropatkin}, {Lampeitl},
  {Lebedeva}, {Lee}, {Lee}, {Leger}, {L{\'e}pine}, {Li}, {Lima}, {Lin}, {Long},
  {Loomis}, {Loveday}, {Lupton}, {Magnier}, {Malanushenko}, {Malanushenko},
  {Mandelbaum}, {Margon}, {Marriner}, {Mart{\'{\i}}nez-Delgado}, {Matsubara},
  {McGehee}, {McKay}, {Meiksin}, {Morrison}, {Mullally}, {Munn}, {Murphy},
  {Nash}, {Nebot}, {Neilsen}, {Newberg}, {Newman}, {Nichol}, {Nicinski},
  {Nieto-Santisteban}, {Nitta}, {Okamura}, {Oravetz}, {Ostriker}, {Owen},
  {Padmanabhan}, {Pan}, {Park}, {Pauls}, {Peoples}, {Percival}, {Pier}, {Pope},
  {Pourbaix}, {Price}, {Purger}, {Quinn}, {Raddick}, {Fiorentin}, {Richards},
  {Richmond}, {Riess}, {Rix}, {Rockosi}, {Sako}, {Schlegel}, {Schneider},
  {Scholz}, {Schreiber}, {Schwope}, {Seljak}, {Sesar}, {Sheldon}, {Shimasaku},
  {Sibley}, {Simmons}, {Sivarani}, {Smith}, {Smith}, {Smol{\v c}i{\'c}},
  {Snedden}, {Stebbins}, {Steinmetz}, {Stoughton}, {Strauss}, {Subba Rao},
  {Suto}, {Szalay}, {Szapudi}, {Szkody}, {Tanaka}, {Tegmark}, {Teodoro},
  {Thakar}, {Tremonti}, {Tucker}, {Uomoto}, {Vanden Berk}, {Vandenberg},
  {Vidrih}, {Vogeley}, {Voges}, {Vogt}, {Wadadekar}, {Watters}, {Weinberg},
  {West}, {White}, {Wilhite}, {Wonders}, {Yanny}, {Yocum}, {York}, {Zehavi},
  {Zibetti}, \& {Zucker}}]{abazajian2009}
{Abazajian}, K.~N., {Adelman-McCarthy}, J.~K., {Ag{\"u}eros}, M.~A., {et~al.}
  2009, \apjs, 182, 543

\bibitem[{{Allen} {et~al.}(2005){Allen}, {Koerner}, {Reid}, \&
  {Trilling}}]{allen2005}
{Allen}, P.~R., {Koerner}, D.~W., {Reid}, I.~N., \& {Trilling}, D.~E. 2005,
  \apj, 625, 385

\bibitem[{{Allers} {et~al.}(2007){Allers}, {Jaffe}, {Luhman}, {Liu}, {Wilson},
  {Skrutskie}, {Nelson}, {Peterson}, {Smith}, \& {Cushing}}]{allers2007}
{Allers}, K.~N., {Jaffe}, D.~T., {Luhman}, K.~L., {et~al.} 2007, \apj, 657, 511

\bibitem[{{Allers} {et~al.}(2010){Allers}, {Liu}, {Dupuy}, \&
  {Cushing}}]{allers2010}
{Allers}, K.~N., {Liu}, M.~C., {Dupuy}, T.~J., \& {Cushing}, M.~C. 2010, \apj,
  715, 561

\bibitem[{{An} {et~al.}(2007){An}, {Terndrup}, \& {Pinsonneault}}]{an2007}
{An}, D., {Terndrup}, D.~M., \& {Pinsonneault}, M.~H. 2007, \apj, 671, 1640

\bibitem[{{Bannister} \& {Jameson}(2007)}]{bannister2007}
{Bannister}, N.~P. \& {Jameson}, R.~F. 2007, \mnras, 378, L24

\bibitem[{{Barrado y Navascu{\'e}s} {et~al.}(2001){Barrado y Navascu{\'e}s},
  {Zapatero Osorio}, {B{\'e}jar}, {Rebolo}, {Mart{\'{\i}}n}, {Mundt}, \&
  {Bailer-Jones}}]{barradoynavascues2001b}
{Barrado y Navascu{\'e}s}, D., {Zapatero Osorio}, M.~R., {B{\'e}jar}, V.~J.~S.,
  {et~al.} 2001, \aap, 377, L9

\bibitem[{{Barrado y Navascu{\'e}s} {et~al.}(2002){Barrado y Navascu{\'e}s},
  {Zapatero Osorio}, {Mart{\'{\i}}n}, {B{\'e}jar}, {Rebolo}, \&
  {Mundt}}]{barradoynavascues2002or}
{Barrado y Navascu{\'e}s}, D., {Zapatero Osorio}, M.~R., {Mart{\'{\i}}n},
  E.~L., {et~al.} 2002, \aap, 393, L85

\bibitem[{{Basri} {et~al.}(1996){Basri}, {Marcy}, \& {Graham}}]{basri1996}
{Basri}, G., {Marcy}, G.~W., \& {Graham}, J.~R. 1996, \apj, 458, 600

\bibitem[{{Basri} {et~al.}(2000){Basri}, {Mohanty}, {Allard}, {Hauschildt},
  {Delfosse}, {Mart{\'{\i}}n}, {Forveille}, \& {Goldman}}]{basri2000b}
{Basri}, G., {Mohanty}, S., {Allard}, F., {et~al.} 2000, \apj, 538, 363

\bibitem[{{B{\'e}jar} {et~al.}(2001){B{\'e}jar}, {Mart{\'{\i}}n}, {Zapatero
  Osorio}, {Rebolo}, {Barrado y Navascu{\' e}s}, {Bailer-Jones}, {Mundt},
  {Baraffe}, {Chabrier}, \& {Allard}}]{bejar2001}
{B{\'e}jar}, V.~J.~S., {Mart{\'{\i}}n}, E.~L., {Zapatero Osorio}, M.~R.,
  {et~al.} 2001, \apj, 556, 830

\bibitem[{{B{\'e}jar} {et~al.}(2008){B{\'e}jar}, {Zapatero Osorio},
  {P{\'e}rez-Garrido}, {{\'A}lvarez}, {Mart{\'{\i}}n}, {Rebolo},
  {Vill{\'o}-P{\'e}rez}, \& {D{\'{\i}}az-S{\'a}nchez}}]{bejar2008}
{B{\'e}jar}, V.~J.~S., {Zapatero Osorio}, M.~R., {P{\'e}rez-Garrido}, A.,
  {et~al.} 2008, \apjl, 673, L185

\bibitem[{{Bihain} {et~al.}(2006){Bihain}, {Rebolo}, {B{\'e}jar}, {Caballero},
  {Bailer-Jones}, {Mundt}, {Acosta-Pulido}, \& {Manchado Torres}}]{bihain2006}
{Bihain}, G., {Rebolo}, R., {B{\'e}jar}, V.~J.~S., {et~al.} 2006, \aap, 458,
  805

\bibitem[{{Bonnefoy} {et~al.}(2010){Bonnefoy}, {Chauvin}, {Rojo}, {Allard},
  {Lagrange}, {Homeier}, {Dumas}, \& {Beuzit}}]{bonnefoy2010}
{Bonnefoy}, M., {Chauvin}, G., {Rojo}, P., {et~al.} 2010, \aap, 512, A52

\bibitem[{{Borysow} {et~al.}(1997){Borysow}, {Jorgensen}, \&
  {Zheng}}]{borysow1997}
{Borysow}, A., {Jorgensen}, U.~G., \& {Zheng}, C. 1997, \aap, 324, 185

\bibitem[{{Bouvier} {et~al.}(1998){Bouvier}, {Stauffer}, {Mart{\'{\i}}n},
  {Barrado y Navascues}, {Wallace}, \& {Bejar}}]{bouvier1998}
{Bouvier}, J., {Stauffer}, J.~R., {Mart{\'{\i}}n}, E.~L., {et~al.} 1998, \aap,
  336, 490

\bibitem[{{Bowler} {et~al.}(2009){Bowler}, {Liu}, \& {Cushing}}]{bowler2009}
{Bowler}, B.~P., {Liu}, M.~C., \& {Cushing}, M.~C. 2009, \apj, 706, 1114

\bibitem[{{Burgasser}(2004)}]{burgasser2004c}
{Burgasser}, A.~J. 2004, \apjs, 155, 191

\bibitem[{{Burgasser}(2007)}]{burgasser2007}
{Burgasser}, A.~J. 2007, \apj, 659, 655

\bibitem[{{Burgasser} {et~al.}(2008){Burgasser}, {Liu}, {Ireland}, {Cruz}, \&
  {Dupuy}}]{burgasser2008}
{Burgasser}, A.~J., {Liu}, M.~C., {Ireland}, M.~J., {Cruz}, K.~L., \& {Dupuy},
  T.~J. 2008, \apj, 681, 579

\bibitem[{{Burgasser} {et~al.}(2004){Burgasser}, {McElwain}, {Kirkpatrick},
  {Cruz}, {Tinney}, \& {Reid}}]{burgasser2004b}
{Burgasser}, A.~J., {McElwain}, M.~W., {Kirkpatrick}, J.~D., {et~al.} 2004,
  \aj, 127, 2856

\bibitem[{{Burrows} {et~al.}(1997){Burrows}, {Marley}, {Hubbard}, {Lunine},
  {Guillot}, {Saumon}, {Freedman}, {Sudarsky}, \& {Sharp}}]{burrows1997}
{Burrows}, A., {Marley}, M., {Hubbard}, W.~B., {et~al.} 1997, \apj, 491, 856

\bibitem[{{Casali} {et~al.}(2007){Casali}, {Adamson}, {Alves de Oliveira},
  {Almaini}, {Burch}, {Chuter}, {Elliot}, {Folger}, {Foucaud}, {Hambly},
  {Hastie}, {Henry}, {Hirst}, {Irwin}, {Ives}, {Lawrence}, {Laidlaw}, {Lee},
  {Lewis}, {Lunney}, {McLay}, {Montgomery}, {Pickup}, {Read}, {Rees}, {Robson},
  {Sekiguchi}, {Vick}, {Warren}, \& {Woodward}}]{casali2007}
{Casali}, M., {Adamson}, A., {Alves de Oliveira}, C., {et~al.} 2007, \aap, 467,
  777

\bibitem[{{Casewell} {et~al.}(2007){Casewell}, {Dobbie}, {Hodgkin}, {Moraux},
  {Jameson}, {Hambly}, {Irwin}, \& {Lodieu}}]{casewell2007}
{Casewell}, S.~L., {Dobbie}, P.~D., {Hodgkin}, S.~T., {et~al.} 2007, \mnras,
  378, 1131

\bibitem[{{Casewell} {et~al.}(2008){Casewell}, {Jameson}, \&
  {Burleigh}}]{casewell2008}
{Casewell}, S.~L., {Jameson}, R.~F., \& {Burleigh}, M.~R. 2008, \mnras, 390,
  1517

\bibitem[{{Chabrier}(2002)}]{chabrier2002}
{Chabrier}, G. 2002, \apj, 567, 304

\bibitem[{{Chabrier}(2003)}]{chabrier2003}
{Chabrier}, G. 2003, \pasp, 115, 763

\bibitem[{{Chabrier} {et~al.}(2000){Chabrier}, {Baraffe}, {Allard}, \&
  {Hauschildt}}]{chabrier2000b}
{Chabrier}, G., {Baraffe}, I., {Allard}, F., \& {Hauschildt}, P. 2000, \apj,
  542, 464

\bibitem[{{Chauvin} {et~al.}(2004){Chauvin}, {Lagrange}, {Dumas}, {Zuckerman},
  {Mouillet}, {Song}, {Beuzit}, \& {Lowrance}}]{chauvin2004}
{Chauvin}, G., {Lagrange}, A.-M., {Dumas}, C., {et~al.} 2004, \aap, 425, L29

\bibitem[{{Chauvin} {et~al.}(2005){Chauvin}, {Lagrange}, {Zuckerman}, {Dumas},
  {Mouillet}, {Song}, {Beuzit}, {Lowrance}, \& {Bessell}}]{chauvin2005b}
{Chauvin}, G., {Lagrange}, A.-M., {Zuckerman}, B., {et~al.} 2005, \aap, 438,
  L29

\bibitem[{{Chiu} {et~al.}(2006){Chiu}, {Fan}, {Leggett}, {Golimowski}, {Zheng},
  {Geballe}, {Schneider}, \& {Brinkmann}}]{chiu2006}
{Chiu}, K., {Fan}, X., {Leggett}, S.~K., {et~al.} 2006, \aj, 131, 2722

\bibitem[{{Cossburn} {et~al.}(1997){Cossburn}, {Hodgkin}, {Jameson}, \&
  {Pinfield}}]{cossburn1997}
{Cossburn}, M.~R., {Hodgkin}, S.~T., {Jameson}, R.~F., \& {Pinfield}, D.~J.
  1997, \mnras, 288, L23

\bibitem[{{Cruz} {et~al.}(2009){Cruz}, {Kirkpatrick}, \&
  {Burgasser}}]{cruz2009}
{Cruz}, K.~L., {Kirkpatrick}, J.~D., \& {Burgasser}, A.~J. 2009, \aj, 137, 3345

\bibitem[{{Cushing} {et~al.}(2005){Cushing}, {Rayner}, \&
  {Vacca}}]{cushing2005}
{Cushing}, M.~C., {Rayner}, J.~T., \& {Vacca}, W.~D. 2005, \apj, 623, 1115

\bibitem[{{Dahn} {et~al.}(2000){Dahn}, {Guetter}, {Harris}, {Henden},
  {Luginbuhl}, {Monet}, {Monet}, {Pier}, {Stone}, {Vrba}, \&
  {Walker}}]{dahn2000}
{Dahn}, C.~C., {Guetter}, H.~H., {Harris}, H.~C., {et~al.} 2000, in ASP Conf.
  Ser. 212: From Giant Planets to Cool Stars, 74

\bibitem[{{Dahn} {et~al.}(2002){Dahn}, {Harris}, {Vrba}, {Guetter}, {Canzian},
  {Henden}, {Levine}, {Luginbuhl}, {Monet}, {Monet}, {Pier}, {Stone}, {Walker},
  {Burgasser}, {Gizis}, {Kirkpatrick}, {Liebert}, \& {Reid}}]{dahn2002}
{Dahn}, C.~C., {Harris}, H.~C., {Vrba}, F.~J., {et~al.} 2002, \aj, 124, 1170

\bibitem[{{Deacon} \& {Hambly}(2004)}]{deacon2004}
{Deacon}, N.~R. \& {Hambly}, N.~C. 2004, \aap, 416, 125

\bibitem[{{Ducourant} {et~al.}(2008){Ducourant}, {Teixeira}, {Chauvin},
  {Daigne}, {Le Campion}, {Song}, \& {Zuckerman}}]{ducourant2008}
{Ducourant}, C., {Teixeira}, R., {Chauvin}, G., {et~al.} 2008, \aap, 477, L1

\bibitem[{{Dupuy} {et~al.}(2009){Dupuy}, {Liu}, \& {Ireland}}]{dupuy2009}
{Dupuy}, T.~J., {Liu}, M.~C., \& {Ireland}, M.~J. 2009, \apj, 692, 729

\bibitem[{{Faherty} {et~al.}(2009){Faherty}, {Burgasser}, {Cruz}, {Shara},
  {Walter}, \& {Gelino}}]{faherty2009}
{Faherty}, J.~K., {Burgasser}, A.~J., {Cruz}, K.~L., {et~al.} 2009, \aj, 137, 1

\bibitem[{{Festin}(1998{\natexlab{a}})}]{festin1998a}
{Festin}, L. 1998{\natexlab{a}}, \aap, 333, 497

\bibitem[{{Festin}(1998{\natexlab{b}})}]{festin1998b}
{Festin}, L. 1998{\natexlab{b}}, \mnras, 298, L34

\bibitem[{{Funayama} {et~al.}(2009){Funayama}, {Itoh}, {Oasa}, {Toyota},
  {Hashimoto}, \& {Mukai}}]{funayama2009}
{Funayama}, H., {Itoh}, Y., {Oasa}, Y., {et~al.} 2009, \pasj, 61, 930

\bibitem[{{Geballe} {et~al.}(2002){Geballe}, {Knapp}, {Leggett}, {Fan},
  {Golimowski}, {Anderson}, {Brinkmann}, {Csabai}, {Gunn}, {Hawley},
  {Hennessy}, {Henry}, {Hill}, {Hindsley}, {Ivezi{\' c}}, {Lupton}, {McDaniel},
  {Munn}, {Narayanan}, {Peng}, {Pier}, {Rockosi}, {Schneider}, {Smith},
  {Strauss}, {Tsvetanov}, {Uomoto}, {York}, \& {Zheng}}]{geballe2002}
{Geballe}, T.~R., {Knapp}, G.~R., {Leggett}, S.~K., {et~al.} 2002, \apj, 564,
  466

\bibitem[{{Gizis}(2002)}]{gizis2002}
{Gizis}, J.~E. 2002, \apj, 575, 484

\bibitem[{{Gizis} {et~al.}(2001){Gizis}, {Kirkpatrick}, \&
  {Wilson}}]{gizis2001}
{Gizis}, J.~E., {Kirkpatrick}, J.~D., \& {Wilson}, J.~C. 2001, \aj, 121, 2185

\bibitem[{{Gizis} {et~al.}(2000){Gizis}, {Monet}, {Reid}, {Kirkpatrick},
  {Liebert}, \& {Williams}}]{gizis2000}
{Gizis}, J.~E., {Monet}, D.~G., {Reid}, I.~N., {et~al.} 2000, \aj, 120, 1085

\bibitem[{{Golimowski} {et~al.}(2004){Golimowski}, {Leggett}, {Marley}, {Fan},
  {Geballe}, {Knapp}, {Vrba}, {Henden}, {Luginbuhl}, {Guetter}, {Munn},
  {Canzian}, {Zheng}, {Tsvetanov}, {Chiu}, {Glazebrook}, {Hoversten},
  {Schneider}, \& {Brinkmann}}]{golimowski2004}
{Golimowski}, D.~A., {Leggett}, S.~K., {Marley}, M.~S., {et~al.} 2004, \aj,
  127, 3516

\bibitem[{{Greissl} {et~al.}(2007){Greissl}, {Meyer}, {Wilking}, {Fanetti},
  {Schneider}, {Greene}, \& {Young}}]{greissl2007}
{Greissl}, J., {Meyer}, M.~R., {Wilking}, B.~A., {et~al.} 2007, \aj, 133, 1321

\bibitem[{{Hambly} {et~al.}(2008){Hambly}, {Collins}, {Cross}, {Mann}, {Read},
  {Sutorius}, {Bond}, {Bryant}, {Emerson}, {Lawrence}, {Rimoldini}, {Stewart},
  {Williams}, {Adamson}, {Hirst}, {Dye}, \& {Warren}}]{hambly2008}
{Hambly}, N.~C., {Collins}, R.~S., {Cross}, N.~J.~G., {et~al.} 2008, \mnras,
  384, 637

\bibitem[{{Hambly} {et~al.}(1993){Hambly}, {Hawkins}, \&
  {Jameson}}]{hambly1993}
{Hambly}, N.~C., {Hawkins}, M.~R.~S., \& {Jameson}, R.~F. 1993, \aaps, 100, 607

\bibitem[{{Hammersley}(1998)}]{hammersley1998}
{Hammersley}, P.~L. 1998, New Astronomy Review, 42, 533

\bibitem[{{Henry} {et~al.}(2004){Henry}, {Subasavage}, {Brown}, {Beaulieu},
  {Jao}, \& {Hambly}}]{henry2004}
{Henry}, T.~J., {Subasavage}, J.~P., {Brown}, M.~A., {et~al.} 2004, \aj, 128,
  2460

\bibitem[{{Hewett} {et~al.}(2006){Hewett}, {Warren}, {Leggett}, \&
  {Hodgkin}}]{hewett2006}
{Hewett}, P.~C., {Warren}, S.~J., {Leggett}, S.~K., \& {Hodgkin}, S.~T. 2006,
  \mnras, 367, 454

\bibitem[{{Hodgkin} {et~al.}(2009){Hodgkin}, {Irwin}, {Hewett}, \&
  {Warren}}]{hodgkin2009}
{Hodgkin}, S.~T., {Irwin}, M.~J., {Hewett}, P.~C., \& {Warren}, S.~J. 2009,
  \mnras, 394, 675

\bibitem[{{Itoh} {et~al.}(2005){Itoh}, {Hayashi}, {Tamura}, {Tsuji}, {Oasa},
  {Fukagawa}, {Hayashi}, {Naoi}, {Ishii}, {Mayama}, {Morino}, {Yamashita},
  {Pyo}, {Nishikawa}, {Usuda}, {Murakawa}, {Suto}, {Oya}, {Takato}, {Ando},
  {Miyama}, {Kobayashi}, \& {Kaifu}}]{itoh2005}
{Itoh}, Y., {Hayashi}, M., {Tamura}, M., {et~al.} 2005, \apj, 620, 984

\bibitem[{{Jameson} {et~al.}(2008){Jameson}, {Casewell}, {Bannister}, {Lodieu},
  {Keresztes}, {Dobbie}, \& {Hodgkin}}]{jameson2008a}
{Jameson}, R.~F., {Casewell}, S.~L., {Bannister}, N.~P., {et~al.} 2008, \mnras,
  384, 1399

\bibitem[{{Jameson} {et~al.}(2002){Jameson}, {Dobbie}, {Hodgkin}, \&
  {Pinfield}}]{jameson2002}
{Jameson}, R.~F., {Dobbie}, P.~D., {Hodgkin}, S.~T., \& {Pinfield}, D.~J. 2002,
  \mnras, 335, 853

\bibitem[{{Jayawardhana} \& {Ivanov}(2006{\natexlab{a}})}]{jayawardhana2006}
{Jayawardhana}, R. \& {Ivanov}, V.~D. 2006{\natexlab{a}}, Science, 313, 1279

\bibitem[{{Jayawardhana} \& {Ivanov}(2006{\natexlab{b}})}]{jayawardhana2006b}
{Jayawardhana}, R. \& {Ivanov}, V.~D. 2006{\natexlab{b}}, \apjl, 647, L167

\bibitem[{{Kirkpatrick} {et~al.}(2006){Kirkpatrick}, {Barman}, {Burgasser},
  {McGovern}, {McLean}, {Tinney}, \& {Lowrance}}]{kirkpatrick2006}
{Kirkpatrick}, J.~D., {Barman}, T.~S., {Burgasser}, A.~J., {et~al.} 2006, \apj,
  639, 1120

\bibitem[{{Kirkpatrick} {et~al.}(2008){Kirkpatrick}, {Cruz}, {Barman},
  {Burgasser}, {Looper}, {Tinney}, {Gelino}, {Lowrance}, {Liebert},
  {Carpenter}, {Hillenbrand}, \& {Stauffer}}]{kirkpatrick2008}
{Kirkpatrick}, J.~D., {Cruz}, K.~L., {Barman}, T.~S., {et~al.} 2008, \apj, 689,
  1295

\bibitem[{{Kirkpatrick} {et~al.}(2001){Kirkpatrick}, {Dahn}, {Monet}, {Reid},
  {Gizis}, {Liebert}, \& {Burgasser}}]{kirkpatrick2001}
{Kirkpatrick}, J.~D., {Dahn}, C.~C., {Monet}, D.~G., {et~al.} 2001, \aj, 121,
  3235

\bibitem[{{Kirkpatrick} {et~al.}(1999){Kirkpatrick}, {Reid}, {Liebert},
  {Cutri}, {Nelson}, {Beichman}, {Dahn}, {Monet}, {Gizis}, \&
  {Skrutskie}}]{kirkpatrick1999b}
{Kirkpatrick}, J.~D., {Reid}, I.~N., {Liebert}, J., {et~al.} 1999, \apj, 519,
  802

\bibitem[{{Kirkpatrick} {et~al.}(2000){Kirkpatrick}, {Reid}, {Liebert},
  {Gizis}, {Burgasser}, {Monet}, {Dahn}, {Nelson}, \&
  {Williams}}]{kirkpatrick2000}
{Kirkpatrick}, J.~D., {Reid}, I.~N., {Liebert}, J., {et~al.} 2000, \aj, 120,
  447

\bibitem[{{Knapp} {et~al.}(2004){Knapp}, {Leggett}, {Fan}, {Marley}, {Geballe},
  {Golimowski}, {Finkbeiner}, {Gunn}, {Hennawi}, {Ivezi{\'c}}, {Lupton},
  {Schlegel}, {Strauss}, {Tsvetanov}, {Chiu}, {Hoversten}, {Glazebrook},
  {Zheng}, {Hendrickson}, {Williams}, {Uomoto}, {Vrba}, {Henden}, {Luginbuhl},
  {Guetter}, {Munn}, {Canzian}, {Schneider}, \& {Brinkmann}}]{knapp2004}
{Knapp}, G.~R., {Leggett}, S.~K., {Fan}, X., {et~al.} 2004, \aj, 127, 3553

\bibitem[{{Kroupa} \& {Bouvier}(2003)}]{kroupa2003}
{Kroupa}, P. \& {Bouvier}, J. 2003, \mnras, 346, 369

\bibitem[{{Lafreni{\`e}re} {et~al.}(2008){Lafreni{\`e}re}, {Jayawardhana}, \&
  {van Kerkwijk}}]{lafreniere2008}
{Lafreni{\`e}re}, D., {Jayawardhana}, R., \& {van Kerkwijk}, M.~H. 2008, \apjl,
  689, L153

\bibitem[{{Lawrence} {et~al.}(2007){Lawrence}, {Warren}, {Almaini}, {Edge},
  {Hambly}, {Jameson}, {Lucas}, {Casali}, {Adamson}, {Dye}, {Emerson},
  {Foucaud}, {Hewett}, {Hirst}, {Hodgkin}, {Irwin}, {Lodieu}, {McMahon},
  {Simpson}, {Smail}, {Mortlock}, \& {Folger}}]{lawrence2007}
{Lawrence}, A., {Warren}, S.~J., {Almaini}, O., {et~al.} 2007, \mnras, 379,
  1599

\bibitem[{{Leggett} {et~al.}(2000){Leggett}, {Allard}, {Dahn}, {Hauschildt},
  {Kerr}, \& {Rayner}}]{leggett2000}
{Leggett}, S.~K., {Allard}, F., {Dahn}, C., {et~al.} 2000, \apj, 535, 965

\bibitem[{{Leggett} {et~al.}(2001){Leggett}, {Allard}, {Geballe}, {Hauschildt},
  \& {Schweitzer}}]{leggett2001}
{Leggett}, S.~K., {Allard}, F., {Geballe}, T.~R., {Hauschildt}, P.~H., \&
  {Schweitzer}, A. 2001, \apj, 548, 908

\bibitem[{{Linsky}(1969)}]{linsky1969}
{Linsky}, J.~L. 1969, \apj, 156, 989

\bibitem[{{Lodieu} {et~al.}(2002){Lodieu}, {Caux}, {Monin}, \&
  {Klotz}}]{lodieu2002}
{Lodieu}, N., {Caux}, E., {Monin}, J.-L., \& {Klotz}, A. 2002, \aap, 383, L15

\bibitem[{{Lodieu} {et~al.}(2007){Lodieu}, {Dobbie}, {Deacon}, {Hodgkin},
  {Hambly}, \& {Jameson}}]{lodieu2007}
{Lodieu}, N., {Dobbie}, P.~D., {Deacon}, N.~R., {et~al.} 2007, \mnras, 380, 712

\bibitem[{{Lodieu} {et~al.}(2008){Lodieu}, {Hambly}, {Jameson}, \&
  {Hodgkin}}]{lodieu2008}
{Lodieu}, N., {Hambly}, N.~C., {Jameson}, R.~F., \& {Hodgkin}, S.~T. 2008,
  \mnras, 383, 1385

\bibitem[{{Lodieu} {et~al.}(2005){Lodieu}, {Scholz}, {McCaughrean}, {Ibata},
  {Irwin}, \& {Zinnecker}}]{lodieu2005}
{Lodieu}, N., {Scholz}, R., {McCaughrean}, M.~J., {et~al.} 2005, \aap, 440,
  1061

\bibitem[{{Looper} {et~al.}(2007){Looper}, {Burgasser}, {Kirkpatrick}, \&
  {Swift}}]{looper2007}
{Looper}, D.~L., {Burgasser}, A.~J., {Kirkpatrick}, J.~D., \& {Swift}, B.~J.
  2007, \apjl, 669, L97

\bibitem[{{Looper} {et~al.}(2008){Looper}, {Kirkpatrick}, {Cutri}, {Barman},
  {Burgasser}, {Cushing}, {Roellig}, {McGovern}, {McLean}, {Rice}, {Swift}, \&
  {Schurr}}]{looper2008}
{Looper}, D.~L., {Kirkpatrick}, J.~D., {Cutri}, R.~M., {et~al.} 2008, \apj,
  686, 528

\bibitem[{{Lucas} {et~al.}(2001){Lucas}, {Roche}, {Allard}, \&
  {Hauschildt}}]{lucas2001}
{Lucas}, P.~W., {Roche}, P.~F., {Allard}, F., \& {Hauschildt}, P.~H. 2001,
  \mnras, 326, 695

\bibitem[{{Lucas} {et~al.}(2006){Lucas}, {Weights}, {Roche}, \&
  {Riddick}}]{lucas2006}
{Lucas}, P.~W., {Weights}, D.~J., {Roche}, P.~F., \& {Riddick}, F.~C. 2006,
  \mnras, 373, L60

\bibitem[{{Luhman} {et~al.}(2008){Luhman}, {Allen}, {Allen}, {Gutermuth},
  {Hartmann}, {Mamajek}, {Megeath}, {Myers}, \& {Fazio}}]{luhman2008b}
{Luhman}, K.~L., {Allen}, L.~E., {Allen}, P.~R., {et~al.} 2008, \apj, 675, 1375

\bibitem[{{Luhman} {et~al.}(2009){Luhman}, {Mamajek}, {Allen}, \&
  {Cruz}}]{luhman2009}
{Luhman}, K.~L., {Mamajek}, E.~E., {Allen}, P.~R., \& {Cruz}, K.~L. 2009, \apj,
  703, 399

\bibitem[{{Luhman} \& {Muench}(2008)}]{luhman2008c}
{Luhman}, K.~L. \& {Muench}, A.~A. 2008, \apj, 684, 654

\bibitem[{{Luhman} {et~al.}(2006){Luhman}, {Wilson}, {Brandner}, {Skrutskie},
  {Nelson}, {Smith}, {Peterson}, {Cushing}, \& {Young}}]{luhman2006}
{Luhman}, K.~L., {Wilson}, J.~C., {Brandner}, W., {et~al.} 2006, \apj, 649, 894

\bibitem[{{Mamajek} \& {Meyer}(2007)}]{mamajek2007}
{Mamajek}, E.~E. \& {Meyer}, M.~R. 2007, \apjl, 668, L175

\bibitem[{{Mart{\'{\i}}n} {et~al.}(1998{\natexlab{a}}){Mart{\'{\i}}n}, {Basri},
  {Gallegos}, {Rebolo}, {Zapatero Osorio}, \& {Bejar}}]{martin1998a}
{Mart{\'{\i}}n}, E.~L., {Basri}, G., {Gallegos}, J.~E., {et~al.}
  1998{\natexlab{a}}, \apjl, 499, L61

\bibitem[{{Mart{\'{\i}}n} {et~al.}(1998{\natexlab{b}}){Mart{\'{\i}}n}, {Basri},
  {Zapatero Osorio}, {Rebolo}, \& {L{\' o}pez}}]{martin1998b}
{Mart{\'{\i}}n}, E.~L., {Basri}, G., {Zapatero Osorio}, M.~R., {Rebolo}, R., \&
  {L{\' o}pez}, R.~J.~G. 1998{\natexlab{b}}, \apjl, 507, L41

\bibitem[{{Mart{\'{\i}}n} {et~al.}(2000){Mart{\'{\i}}n}, {Brandner}, {Bouvier},
  {Luhman}, {Stauffer}, {Basri}, {Zapatero Osorio}, \& {Barrado y Navascu{\'
  e}s}}]{martin2000}
{Mart{\'{\i}}n}, E.~L., {Brandner}, W., {Bouvier}, J., {et~al.} 2000, \apj,
  543, 299

\bibitem[{{Mart{\'{\i}}n} {et~al.}(1999){Mart{\'{\i}}n}, {Delfosse}, {Basri},
  {Goldman}, {Forveille}, \& {Zapatero Osorio}}]{martin1999}
{Mart{\'{\i}}n}, E.~L., {Delfosse}, X., {Basri}, G., {et~al.} 1999, \aj, 118,
  2466

\bibitem[{{Mart{\'{\i}}n} {et~al.}(1996){Mart{\'{\i}}n}, {Rebolo}, \& {Zapatero
  Osorio}}]{martin1996}
{Mart{\'{\i}}n}, E.~L., {Rebolo}, R., \& {Zapatero Osorio}, M.~R. 1996, \apj,
  469, 706

\bibitem[{{Mart{\'{\i}}n} {et~al.}(2001){Mart{\'{\i}}n}, {Zapatero Osorio},
  {Barrado y Navascu{\'e}s}, {B{\'e}jar}, \& {Rebolo}}]{martin2001}
{Mart{\'{\i}}n}, E.~L., {Zapatero Osorio}, M.~R., {Barrado y Navascu{\'e}s},
  D., {B{\'e}jar}, V.~J.~S., \& {Rebolo}, R. 2001, \apjl, 558, L117

\bibitem[{{McElwain} {et~al.}(2007){McElwain}, {Metchev}, {Larkin}, {Barczys},
  {Iserlohe}, {Krabbe}, {Quirrenbach}, {Weiss}, \& {Wright}}]{mcelwain2007}
{McElwain}, M.~W., {Metchev}, S.~A., {Larkin}, J.~E., {et~al.} 2007, \apj, 656,
  505

\bibitem[{{McGovern} {et~al.}(2004){McGovern}, {Kirkpatrick}, {McLean},
  {Burgasser}, {Prato}, \& {Lowrance}}]{mcgovern2004}
{McGovern}, M.~R., {Kirkpatrick}, J.~D., {McLean}, I.~S., {et~al.} 2004, \apj,
  600, 1020

\bibitem[{{Metchev} \& {Hillenbrand}(2004)}]{metchev2004}
{Metchev}, S.~A. \& {Hillenbrand}, L.~A. 2004, \apj, 617, 1330

\bibitem[{{Mohanty} {et~al.}(2007){Mohanty}, {Jayawardhana}, {Hu{\'e}lamo}, \&
  {Mamajek}}]{mohanty2007}
{Mohanty}, S., {Jayawardhana}, R., {Hu{\'e}lamo}, N., \& {Mamajek}, E. 2007,
  \apj, 657, 1064

\bibitem[{{Moraux} {et~al.}(2001){Moraux}, {Bouvier}, \&
  {Stauffer}}]{moraux2001}
{Moraux}, E., {Bouvier}, J., \& {Stauffer}, J.~R. 2001, \aap, 367, 211

\bibitem[{{Moraux} {et~al.}(2003){Moraux}, {Bouvier}, {Stauffer}, \&
  {Cuillandre}}]{moraux2003}
{Moraux}, E., {Bouvier}, J., {Stauffer}, J.~R., \& {Cuillandre}, J.-C. 2003,
  \aap, 400, 891

\bibitem[{{Neuh{\"a}user} {et~al.}(2005){Neuh{\"a}user}, {Guenther},
  {Wuchterl}, {Mugrauer}, {Bedalov}, \& {Hauschildt}}]{neuhauser2005}
{Neuh{\"a}user}, R., {Guenther}, E.~W., {Wuchterl}, G., {et~al.} 2005, \aap,
  435, L13

\bibitem[{{Patience} {et~al.}(2010){Patience}, {King}, {De Rosa}, \&
  {Marois}}]{patience2010}
{Patience}, J., {King}, R.~R., {De Rosa}, R.~J., \& {Marois}, C. 2010, \aap~in
  press, DOI: 10.1051/0004-6361/201014173

\bibitem[{{Percival} {et~al.}(2005){Percival}, {Salaris}, \&
  {Groenewegen}}]{percival2005}
{Percival}, S.~M., {Salaris}, M., \& {Groenewegen}, M.~A.~T. 2005, \aap, 429,
  887

\bibitem[{{Phan-Bao} {et~al.}(2008){Phan-Bao}, {Bessell}, {Mart{\'{\i}}n},
  {Simon}, {Borsenberger}, {Tata}, {Guibert}, {Crifo}, {Forveille}, {Delfosse},
  {Lim}, \& {de Batz}}]{phan-bao2008}
{Phan-Bao}, N., {Bessell}, M.~S., {Mart{\'{\i}}n}, E.~L., {et~al.} 2008,
  \mnras, 383, 831

\bibitem[{{Pinfield} {et~al.}(2003){Pinfield}, {Dobbie}, {Jameson}, {Steele},
  {Jones}, \& {Katsiyannis}}]{pinfield2003}
{Pinfield}, D.~J., {Dobbie}, P.~D., {Jameson}, R.~F., {et~al.} 2003, \mnras,
  342, 1241

\bibitem[{{Pinfield} {et~al.}(2000){Pinfield}, {Hodgkin}, {Jameson},
  {Cossburn}, {Hambly}, \& {Devereux}}]{pinfield2000}
{Pinfield}, D.~J., {Hodgkin}, S.~T., {Jameson}, R.~F., {et~al.} 2000, \mnras,
  313, 347

\bibitem[{{Pinfield} {et~al.}(1998){Pinfield}, {Jameson}, \&
  {Hodgkin}}]{pinfield1998}
{Pinfield}, D.~J., {Jameson}, R.~F., \& {Hodgkin}, S.~T. 1998, \mnras, 299, 955

\bibitem[{{Pinfield} {et~al.}(2006){Pinfield}, {Jones}, {Lucas}, {Kendall},
  {Folkes}, {Day-Jones}, {Chappelle}, \& {Steele}}]{pinfield2006}
{Pinfield}, D.~J., {Jones}, H.~R.~A., {Lucas}, P.~W., {et~al.} 2006, \mnras,
  368, 1281

\bibitem[{{Potter} {et~al.}(2002){Potter}, {Mart{\'{\i}}n}, {Cushing},
  {Baudoz}, {Brandner}, {Guyon}, \& {Neuh{\"a}user}}]{potter2002}
{Potter}, D., {Mart{\'{\i}}n}, E.~L., {Cushing}, M.~C., {et~al.} 2002, \apjl,
  567, L133

\bibitem[{{Rayner} {et~al.}(2009){Rayner}, {Cushing}, \& {Vacca}}]{rayner2009}
{Rayner}, J.~T., {Cushing}, M.~C., \& {Vacca}, W.~D. 2009, \apjs, 185, 289

\bibitem[{{Rebolo} {et~al.}(1996){Rebolo}, {Mart{\'{\i}}n}, {Basri}, {Marcy},
  \& {Zapatero Osorio}}]{rebolo1996}
{Rebolo}, R., {Mart{\'{\i}}n}, E.~L., {Basri}, G., {Marcy}, G.~W., \& {Zapatero
  Osorio}, M.~R. 1996, \apjl, 469, L53

\bibitem[{{Rebolo} {et~al.}(1992){Rebolo}, {Mart{\'{\i}}n}, \&
  {Magazz\`u}}]{rebolo1992}
{Rebolo}, R., {Mart{\'{\i}}n}, E.~L., \& {Magazz\`u}, A. 1992, \apjl, 389, L83

\bibitem[{{Rebolo} {et~al.}(1998){Rebolo}, {Zapatero Osorio}, {Madruga},
  {Bejar}, {Arribas}, \& {Licandro}}]{rebolo1998}
{Rebolo}, R., {Zapatero Osorio}, M.~R., {Madruga}, S., {et~al.} 1998, Science,
  282, 1309

\bibitem[{{Reid} {et~al.}(2001){Reid}, {Burgasser}, {Cruz}, {Kirkpatrick}, \&
  {Gizis}}]{reid2001b}
{Reid}, I.~N., {Burgasser}, A.~J., {Cruz}, K.~L., {Kirkpatrick}, J.~D., \&
  {Gizis}, J.~E. 2001, \aj, 121, 1710

\bibitem[{{Reid} {et~al.}(2008){Reid}, {Cruz}, {Kirkpatrick}, {Allen},
  {Mungall}, {Liebert}, {Lowrance}, \& {Sweet}}]{reid2008}
{Reid}, I.~N., {Cruz}, K.~L., {Kirkpatrick}, J.~D., {et~al.} 2008, \aj, 136,
  1290

\bibitem[{{Reid} {et~al.}(1999){Reid}, {Kirkpatrick}, {Liebert}, {Burrows},
  {Gizis}, {Burgasser}, {Dahn}, {Monet}, {Cutri}, {Beichman}, \&
  {Skrutskie}}]{reid1999}
{Reid}, I.~N., {Kirkpatrick}, J.~D., {Liebert}, J., {et~al.} 1999, \apj, 521,
  613

\bibitem[{{Ribas}(2003)}]{ribas2003}
{Ribas}, I. 2003, \aap, 400, 297

\bibitem[{{Robichon} {et~al.}(1999){Robichon}, {Arenou}, {Mermilliod}, \&
  {Turon}}]{robichon1999}
{Robichon}, N., {Arenou}, F., {Mermilliod}, J.-C., \& {Turon}, C. 1999, \aap,
  345, 471

\bibitem[{{Saumon} {et~al.}(1994){Saumon}, {Bergeron}, {Lunine}, {Hubbard}, \&
  {Burrows}}]{saumon1994}
{Saumon}, D., {Bergeron}, P., {Lunine}, J.~I., {Hubbard}, W.~B., \& {Burrows},
  A. 1994, \apj, 424, 333

\bibitem[{{Schmidt} {et~al.}(2010){Schmidt}, {West}, {Hawley}, \&
  {Pineda}}]{schmidt2010}
{Schmidt}, S.~J., {West}, A.~A., {Hawley}, S.~L., \& {Pineda}, J.~S. 2010, \aj,
  139, 1808

\bibitem[{{Schmidt} {et~al.}(2008){Schmidt}, {Neuh{\"a}user}, {Seifahrt},
  {Vogt}, {Bedalov}, {Helling}, {Witte}, \& {Hauschildt}}]{schmidt2008}
{Schmidt}, T.~O.~B., {Neuh{\"a}user}, R., {Seifahrt}, A., {et~al.} 2008, \aap,
  491, 311

\bibitem[{{Scholz} \& {Meusinger}(2002)}]{scholz2002}
{Scholz}, R.-D. \& {Meusinger}, H. 2002, \mnras, 336, L49

\bibitem[{{Skrutskie} {et~al.}(2006){Skrutskie}, {Cutri}, {Stiening},
  {Weinberg}, {Schneider}, {Carpenter}, {Beichman}, {Capps}, {Chester},
  {Elias}, {Huchra}, {Liebert}, {Lonsdale}, {Monet}, {Price}, {Seitzer},
  {Jarrett}, {Kirkpatrick}, {Gizis}, {Howard}, {Evans}, {Fowler}, {Fullmer},
  {Hurt}, {Light}, {Kopan}, {Marsh}, {McCallon}, {Tam}, {Van Dyk}, \&
  {Wheelock}}]{skrutskie2006}
{Skrutskie}, M.~F., {Cutri}, R.~M., {Stiening}, R., {et~al.} 2006, \aj, 131,
  1163

\bibitem[{{Slesnick} {et~al.}(2004){Slesnick}, {Hillenbrand}, \&
  {Carpenter}}]{slesnick2004}
{Slesnick}, C.~L., {Hillenbrand}, L.~A., \& {Carpenter}, J.~M. 2004, \apj, 610,
  1045

\bibitem[{{Soderblom} {et~al.}(2009){Soderblom}, {Laskar}, {Valenti},
  {Stauffer}, \& {Rebull}}]{soderblom2009}
{Soderblom}, D.~R., {Laskar}, T., {Valenti}, J.~A., {Stauffer}, J.~R., \&
  {Rebull}, L.~M. 2009, \aj, 138, 1292

\bibitem[{{Stauffer} {et~al.}(1994){Stauffer}, {Hamilton}, \&
  {Probst}}]{stauffer1994}
{Stauffer}, J.~R., {Hamilton}, D., \& {Probst}, R.~G. 1994, \aj, 108, 155

\bibitem[{{Stauffer} {et~al.}(2007){Stauffer}, {Hartmann}, {Fazio}, {Allen},
  {Patten}, {Lowrance}, {Hurt}, {Rebull}, {Cutri}, {Ramirez}, {Young}, {Rieke},
  {Gorlova}, {Muzerolle}, {Slesnick}, \& {Skrutskie}}]{stauffer2007}
{Stauffer}, J.~R., {Hartmann}, L.~W., {Fazio}, G.~G., {et~al.} 2007, \apjs,
  172, 663

\bibitem[{{Stauffer} {et~al.}(1998{\natexlab{a}}){Stauffer}, {Schild}, {Barrado
  y Navascu\'es}, {Backman}, {Angelova}, {Kirkpatrick}, {Hambly}, \&
  {Vanzi}}]{stauffer1998b}
{Stauffer}, J.~R., {Schild}, R., {Barrado y Navascu\'es}, D., {et~al.}
  1998{\natexlab{a}}, \apj, 504, 805

\bibitem[{{Stauffer} {et~al.}(1998{\natexlab{b}}){Stauffer}, {Schultz}, \&
  {Kirkpatrick}}]{stauffer1998a}
{Stauffer}, J.~R., {Schultz}, G., \& {Kirkpatrick}, J.~D. 1998{\natexlab{b}},
  \apjl, 499, L199

\bibitem[{{Steele} \& {Jameson}(1995)}]{steele1995a}
{Steele}, I.~A. \& {Jameson}, R.~F. 1995, \mnras, 272, 630

\bibitem[{{Steele} {et~al.}(1995){Steele}, {Jameson}, {Hodgkin}, \&
  {Hambly}}]{steele1995b}
{Steele}, I.~A., {Jameson}, R.~F., {Hodgkin}, S.~T., \& {Hambly}, N.~C. 1995,
  \mnras, 275, 841

\bibitem[{{Taylor}(2008)}]{taylor2008}
{Taylor}, B.~J. 2008, \aj, 136, 1388

\bibitem[{{Testi}(2009)}]{testi2009}
{Testi}, L. 2009, \aap, 503, 639

\bibitem[{{Testi} {et~al.}(2001){Testi}, {D'Antona}, {Ghinassi}, {Licandro},
  {Magazz{\` u}}, {Maiolino}, {Mannucci}, {Marconi}, {Nagar}, {Natta}, \&
  {Oliva}}]{testi2001}
{Testi}, L., {D'Antona}, F., {Ghinassi}, F., {et~al.} 2001, \apjl, 552, L147

\bibitem[{{Tinney}(1998)}]{tinney1998}
{Tinney}, C.~G. 1998, \mnras, 296, L42

\bibitem[{{Tsuji} {et~al.}(1996){Tsuji}, {Ohnaka}, \& {Aoki}}]{tsuji1996a}
{Tsuji}, T., {Ohnaka}, K., \& {Aoki}, W. 1996, \aap, 305, L1

\bibitem[{{van Altena} {et~al.}(1995){van Altena}, {Lee}, \&
  {Hoffleit}}]{vanaltena1995}
{van Altena}, W.~F., {Lee}, J.~T., \& {Hoffleit}, E.~D. 1995, {The general
  catalogue of trigonometric [stellar] parallaxes}, ed. L.~J. T. . H. E.~D. van
  Altena, W.~F.

\bibitem[{{van Leeuwen}(2009{\natexlab{a}})}]{vanleeuwen2009}
{van Leeuwen}, F. 2009{\natexlab{a}}, \aap, 497, 209

\bibitem[{{van Leeuwen}(2009{\natexlab{b}})}]{vanleeuwen2009b}
{van Leeuwen}, F. 2009{\natexlab{b}}, \aap, 500, 505

\bibitem[{{Vrba} {et~al.}(2004){Vrba}, {Henden}, {Luginbuhl}, {Guetter},
  {Munn}, {Canzian}, {Burgasser}, {Kirkpatrick}, {Fan}, {Geballe},
  {Golimowski}, {Knapp}, {Leggett}, {Schneider}, \& {Brinkmann}}]{vrba2004}
{Vrba}, F.~J., {Henden}, A.~A., {Luginbuhl}, C.~B., {et~al.} 2004, \aj, 127,
  2948

\bibitem[{{Weights} {et~al.}(2009){Weights}, {Lucas}, {Roche}, {Pinfield}, \&
  {Riddick}}]{weights2009}
{Weights}, D.~J., {Lucas}, P.~W., {Roche}, P.~F., {Pinfield}, D.~J., \&
  {Riddick}, F. 2009, \mnras, 392, 817

\bibitem[{{Williams} {et~al.}(1996){Williams}, {Boyle}, {Morgan}, {Rieke},
  {Stauffer}, \& {Rieke}}]{williams1996}
{Williams}, D.~M., {Boyle}, R.~P., {Morgan}, W.~T., {et~al.} 1996, \apj, 464,
  238

\bibitem[{{Zapatero Osorio} {et~al.}(2000){Zapatero Osorio}, {B{\' e}jar},
  {Mart{\'{\i}}n}, {Rebolo}, {Barrado y Navascu{\' e}s}, {Bailer-Jones}, \&
  {Mundt}}]{zapateroosorio2000}
{Zapatero Osorio}, M.~R., {B{\' e}jar}, V.~J.~S., {Mart{\'{\i}}n}, E.~L.,
  {et~al.} 2000, Science, 290, 103

\bibitem[{{Zapatero Osorio} {et~al.}(1999{\natexlab{a}}){Zapatero Osorio},
  {B{\' e}jar}, {Rebolo}, {Mart{\'{\i}}n}, \& {Basri}}]{zapateroosorio1999Or}
{Zapatero Osorio}, M.~R., {B{\' e}jar}, V.~J.~S., {Rebolo}, R.,
  {Mart{\'{\i}}n}, E.~L., \& {Basri}, G. 1999{\natexlab{a}}, \apjl, 524, L115

\bibitem[{{Zapatero Osorio} {et~al.}(1997){Zapatero Osorio}, {Rebolo},
  {Mart{\'{\i}}n}, {Basri}, {Magazzu}, {Hodgkin}, {Jameson}, \&
  {Cossburn}}]{zapateroosorio1997b}
{Zapatero Osorio}, M.~R., {Rebolo}, R., {Mart{\'{\i}}n}, E.~L., {et~al.} 1997,
  \apjl, 491, L81

\bibitem[{{Zapatero Osorio} {et~al.}(1999{\natexlab{b}}){Zapatero Osorio},
  {Rebolo}, {Mart{\'{\i}}n}, {Hodgkin}, {Cossburn}, {Magazz{\` u}}, {Steele},
  \& {Jameson}}]{zapateroosorio1999Pl}
{Zapatero Osorio}, M.~R., {Rebolo}, R., {Mart{\'{\i}}n}, E.~L., {et~al.}
  1999{\natexlab{b}}, \aaps, 134, 537

\end{thebibliography}

\appendix

\section{Pleiades spectroscopic sample of low-mass stars and brown dwarfs}

\begin{table*}
\caption{Pleiades spectroscopic sample of 45 low-mass stars and brown dwarfs.}         
\label{sample}      
\centering          
     {\small	
\begin{tabular}{l c c c c c c c c c}
\hline\hline       
Object & SpT & $\alpha$\tablefootmark{a}\,\,\,\,\,\,\,\,\,\,\,\,\,\,\,\,\,\,\,\,\,\,\,~$\delta$\tablefootmark{a} & $I_{\rm C}$ & $Z$\tablefootmark{b} & $Y$\tablefootmark{b} & $J$\tablefootmark{b} & $H$\tablefootmark{b} & $K$\tablefootmark{b} & Ref.\tablefootmark{c} \\ 
       &     &                                                                    &  (mag)      &(mag)&(mag)&(mag)&(mag)&(mag)& \\ 
\hline	 
\object{HHJ~2}           &  M6.5 & 03 38 27.52~~~25 30 18.1 &  17.30 &  16.591$\pm$0.011   &  15.938$\pm$0.007   &  15.284$\pm$0.007   &  14.747$\pm$0.005   &  14.355$\pm$0.006   & 1,--,1,7  \\ 
\object{HHJ~3}           &  M6.0 & 03 48 50.45~~~22 44 29.9 &  17.51 &  16.562$\pm$0.010   &  15.825$\pm$0.006   &  15.098$\pm$0.006   &  14.531$\pm$0.005   &  14.141$\pm$0.004   & 1,--,1,7  \\ 
\object{HHJ~6}           &  M6.5 & 03 41 42.41~~~23 54 57.1 &  17.00 &  16.171$\pm$0.007   &  15.464$\pm$0.006   &  14.709$\pm$0.004   &  14.124$\pm$0.004   &  13.760$\pm$0.050   & 1,--,1,7  \\ 
\object{HHJ~7}           &  M6.0 & 03 57 49.37~~~22 08 31.0 &  17.00 &  16.149$\pm$0.007   &  15.498$\pm$0.005   &  14.831$\pm$0.005   &  14.286$\pm$0.005   &  13.884$\pm$0.003   & 1,--,1,7  \\ 
\object{PPl~1}           &  M6.5 & 03 45 41.27~~~23 54 09.8 &  18.21 &  17.166$\pm$0.014   &  16.189$\pm$0.008   &  15.360$\pm$0.006   &  14.782$\pm$0.005   &  14.305$\pm$0.006   & 4,8,7,5   \\ 
\object{PPl~14}          &  M5.5 & 03 44 34.30~~~23 51 24.6 &  17.43 &  16.920$\pm$0.011   &  16.157$\pm$0.007   &  15.438$\pm$0.007   &  14.884$\pm$0.005   &  14.478$\pm$0.006   & 4,2,7,1   \\ 
\object{PPl~15}          &  M6.5 & 03 48 04.66~~~23 39 30.2 &  17.91 &  17.014$\pm$0.012   &  16.054$\pm$0.007   &  15.283$\pm$0.006   &  14.704$\pm$0.005   &  14.256$\pm$0.006   & 4,2,7,2   \\ 
\object{Calar~3}         &  M8.0 & 03 51 25.57~~~23 45 21.3 &  19.00 &         ...         &         ...         &  16.083$\pm$0.010   &  15.478$\pm$0.009   &  15.009$\pm$0.011   & 4,3,5,3   \\ 
\object{Teide~1}         &  M8.0 & 03 47 17.92~~~24 22 31.7 &  19.26 &  18.174$\pm$0.029   &  17.067$\pm$0.013   &  16.215$\pm$0.010   &  15.591$\pm$0.009   &  15.096$\pm$0.011   & 4,3,5,2   \\ 
\object{PIZ~1}           &  M9.0 & 03 48 31.52~~~24 34 37.3 &  20.00 &  19.218$\pm$0.065   &  17.883$\pm$0.025   &  16.715$\pm$0.015   &  15.977$\pm$0.012   &  15.357$\pm$0.014   & 5,--,7,6  \\ 
\object{Roque~4}\tablefootmark{d}   &  M9.0 & 03 43 53.55~~~24 31 11.6 &  20.25 &  18.970$\pm$0.054   &  17.709$\pm$0.025   &  16.649$\pm$0.016   &  15.942$\pm$0.016   &  15.260$\pm$0.040   & 6,--,--,5 \\ 
\object{Roque~11}        &  M8.0 & 03 47 12.08~~~24 28 31.7 &  19.06 &         ...         &  16.988$\pm$0.013   &  16.174$\pm$0.087   &  15.505$\pm$0.009   &  15.048$\pm$0.012   & 6,--,8,2  \\ 
\object{Roque~13}        &  M7.5 & 03 45 50.66~~~24 09 03.5 &  18.67 &  17.478$\pm$0.017   &  16.582$\pm$0.010   &  15.705$\pm$0.008   &  15.095$\pm$0.006   &  14.580$\pm$0.008   & 6,8,7,5   \\ 
\object{Roque~14}        &  M7.0 & 03 46 42.98~~~24 24 50.7 &  18.27 &  17.484$\pm$0.018   &  16.403$\pm$0.009   &  15.550$\pm$0.066   &  14.890$\pm$0.006   &  14.391$\pm$0.007   & 6,--,8,7  \\ 
\object{Roque~16}        &  M6.0 & 03 47 39.02~~~24 36 22.2 &  17.91 &  17.112$\pm$0.013   &  16.271$\pm$0.008   &  15.548$\pm$0.007   &  14.992$\pm$0.006   &  14.570$\pm$0.008   & 6,8,7,3   \\ 
\object{Roque~17}        &  M6.5 & 03 47 23.97~~~22 42 37.4 &  17.87 &         ...         &  16.123$\pm$0.008   &  15.354$\pm$0.007   &  14.809$\pm$0.006   &  14.402$\pm$0.007   & 6,--,2,7  \\ 
\object{Teide~2}         &  M6.0 & 03 52 06.72~~~24 16 00.5 &  18.02 &  17.079$\pm$0.013   &  16.255$\pm$0.009   &  15.515$\pm$0.008   &  14.971$\pm$0.008   &  14.507$\pm$0.008   & 7,7,7,3   \\ 
\object{CFHT~9}          &  M6.5 & 03 49 15.12~~~24 36 22.5 &  17.71 &  16.847$\pm$0.011   &  16.059$\pm$0.007   &  15.371$\pm$0.006   &  14.890$\pm$0.005   &  14.433$\pm$0.007   & 8,8,7,3   \\ 
\object{CFHT~10}         &  M6.5 & 03 44 32.32~~~25 25 18.0 &  17.78 &  16.994$\pm$0.012   &  16.209$\pm$0.009   &  15.450$\pm$0.008   &  14.883$\pm$0.007   &  14.476$\pm$0.007   & 8,8,7,3   \\ 
\object{CFHT~12}         &  M8.0 & 03 53 55.12~~~23 23 36.2 &  17.97 &  16.947$\pm$0.012   &  16.027$\pm$0.007   &  15.172$\pm$0.006   &  14.569$\pm$0.005   &  14.088$\pm$0.005   & 8,8,3,3   \\ 
\object{CFHT~15}         &  M7.0 & 03 55 12.60~~~23 17 37.3 &  18.62 &  17.842$\pm$0.022   &  16.877$\pm$0.012   &  15.977$\pm$0.010   &  15.348$\pm$0.009   &  14.842$\pm$0.010   & 8,8,3,3   \\ 
\object{MHObd~3 }        &  M8.0 & 03 41 54.16~~~23 05 04.7 &  18.27 &  17.349$\pm$0.016   &  16.376$\pm$0.010   &  15.522$\pm$0.008   &  14.975$\pm$0.007   &  14.415$\pm$0.007   & 8,8,7,4   \\ 
\object{NPL~26}          &  M5.0 & 03 47 07.89~~~24 23 37.9 &  15.70 &  15.094$\pm$0.004   &  14.598$\pm$0.003   &  13.954$\pm$0.003   &  13.415$\pm$0.002   &  13.080$\pm$0.002   & 9,--,4,2 \\ 
\object{NPL~34}          &  M6.0 & 03 48 55.65~~~24 21 40.1 &  17.05 &  16.220$\pm$0.007   &  15.636$\pm$0.005   &  14.963$\pm$0.005   &  14.416$\pm$0.004   &  14.055$\pm$0.005   & 9,--,1,2 \\ 
\object{NPL~36}          &  M7.5 & 03 48 19.02~~~24 25 12.7 &  18.66 &  17.655$\pm$0.020   &  16.668$\pm$0.011   &  15.930$\pm$0.009   &  15.365$\pm$0.008   &  14.935$\pm$0.011   & 9,--,7,2 \\ 
\object{NPL~38}          &  M8.0 & 03 47 50.41~~~23 54 47.9 &  19.18 &  18.179$\pm$0.028   &  17.138$\pm$0.014   &  16.311$\pm$0.011   &  15.622$\pm$0.009   &  15.093$\pm$0.012   & 9,--,8,2 \\ 
\object{MHObd~1}         &  M7.0 & 03 44 52.42~~~24 36 49.5 &  17.95 &  17.147$\pm$0.014   &  16.294$\pm$0.008   &  15.487$\pm$0.059   &  14.948$\pm$0.006   &  14.525$\pm$0.007   & 10,10,--,4 \\ 
\object{Roque~5}         &  M9.0 & 03 44 22.44~~~23 39 01.3 &  20.19 &  19.107$\pm$0.064   &  17.857$\pm$0.025   &  16.874$\pm$0.017   &  16.254$\pm$0.016   &  15.666$\pm$0.018   & 11,--,7,5 \\ 
\object{CFHT~1}          &  M4.9 & 03 51 51.56~~~23 34 49.1 &  16.10 &  15.431$\pm$0.005   &  14.857$\pm$0.004   &  14.260$\pm$0.003   &  13.768$\pm$0.003   &  13.385$\pm$0.003   & 12,--,3,3 \\ 
\object{CFHT~2}          &  M4.9 & 03 52 44.29~~~23 54 15.0 &  16.61 &  15.738$\pm$0.005   &  15.184$\pm$0.004   &  14.554$\pm$0.004   &  14.000$\pm$0.004   &  13.654$\pm$0.004   & 12,--,3,3 \\ 
\object{CFHT~5}          &  M5.5 & 03 48 44.69~~~24 37 23.5 &  17.02 &  16.260$\pm$0.008   &  15.584$\pm$0.005   &  14.911$\pm$0.005   &  14.419$\pm$0.004   &  14.002$\pm$0.005   & 12,--,7,3 \\ 
\object{CFHT~7}          &  M5.6 & 03 52 05.83~~~24 17 31.0 &  17.34 &  16.512$\pm$0.009   &  15.860$\pm$0.007   &  15.176$\pm$0.006   &  14.618$\pm$0.006   &  14.251$\pm$0.007   & 12,--,7,3 \\ 
\object{CFHT~16}         &  M9.3 & 03 44 35.16~~~25 13 42.8 &  18.66 &  17.656$\pm$0.020   &  16.584$\pm$0.010   &  15.662$\pm$0.007   &  14.985$\pm$0.006   &  14.448$\pm$0.008   & 12,12,7,3 \\ 
\object{CFHT~17}         &  M7.9 & 03 43 00.17~~~24 43 52.3 &  18.80 &  17.791$\pm$0.021   &  16.841$\pm$0.013   &  16.022$\pm$0.010   &  15.425$\pm$0.010   &  15.070$\pm$0.060   & 12,--,3,3 \\ 
\object{CFHT~25}         &  M9.0 & 03 54 05.35~~~23 33 59.3 &  19.69 &  18.651$\pm$0.041   &  17.573$\pm$0.024   &  16.647$\pm$0.018   &  15.964$\pm$0.016   &  15.434$\pm$0.017   & 12,--,3,3 \\ 
\object{NPL~40}          &  M9.8 & 03 48 49.03~~~24 20 25.4 &  20.55 &  19.222$\pm$0.069   &  18.227$\pm$0.035   &  17.237$\pm$0.023   &  16.569$\pm$0.020   &  16.059$\pm$0.029   & 12,--,8,2 \\ 
\object{Roque~7}         &  M8.3 & 03 43 40.31~~~24 30 11.2 &  19.50 &  18.552$\pm$0.038   &  17.490$\pm$0.021   &  16.494$\pm$0.014   &  15.853$\pm$0.015   &  15.470$\pm$0.020   & 12,--,3,3 \\ 
\object{Roque~25}        &  L0.1 & 03 48 30.75~~~22 44 50.4 &  21.80 &  20.389$\pm$0.193   &  18.936$\pm$0.065   &  17.714$\pm$0.039   &  16.861$\pm$0.035   &  16.251$\pm$0.027   & 12,--,9,5 \\ 
\object{BPL~327}         &  M7.1 & 03 55 23.08~~~24 49 04.9 &  18.07 &  17.087$\pm$0.014   &  16.311$\pm$0.009   &  15.528$\pm$0.008   &  15.013$\pm$0.007   &  14.595$\pm$0.008   & 13,--,6,5 \\ 
\object{BRB~17}          &  L0.0 & 03 54 07.98~~~23 54 27.9 &  20.92 &  19.841$\pm$0.111   &  18.451$\pm$0.050   &  17.408$\pm$0.033   &  16.628$\pm$0.030   &  16.028$\pm$0.029   & 14,--,5,6  \\ 
\object{PLIZ~28}         &  L0.0 & 03 54 14.06~~~23 17 52.0 &  21.20 &  20.028$\pm$0.136   &  18.873$\pm$0.062   &  17.601$\pm$0.035   &  16.818$\pm$0.031   &  16.138$\pm$0.030   & 14,--,5,6  \\ 
\object{PLIZ~35}         &  L2.0 & 03 52 39.16~~~24 46 29.5 &  21.46 &  20.292$\pm$0.030   &  19.271$\pm$0.100   &  18.066$\pm$0.055   &  17.106$\pm$0.042   &  16.509$\pm$0.042   & 14,--,5,6  \\ 
\object{BRB~21}          &  L3.0 & 03 54 10.27~~~23 41 40.2 &  21.68 &  20.215$\pm$0.030   &  19.260$\pm$0.101   &  18.143$\pm$0.062   &  17.171$\pm$0.046   &  16.393$\pm$0.040   & 14,--,5,6  \\ 
\object{BRB~23}          &  L3.5 & 03 50 39.54~~~25 02 54.7 &  22.03 &         ...         &  19.786$\pm$0.112   &  18.225$\pm$0.044   &  17.358$\pm$0.038   &  16.563$\pm$0.043   & 14,--,5,6  \\ 
\object{BRB~29}          &  L4.5 & 03 54 01.43~~~23 49 57.7 &  22.84 &         ...         &         ...         &  18.686$\pm$0.103   &  17.711$\pm$0.081   &  16.999$\pm$0.070   & 14,--,5,6  \\ 
\hline                  
\end{tabular}
\begin{flushleft}
\tablefoottext{a}{Coordinates from UKIDSS Galaxy Cluster Survey DDR6.}\\
\tablefoottext{b}{$ZYJHK$-band photometry from UKIDSS Galaxy Cluster Survey DDR6, except for the
measurements at $Z$~band of PLIZ~35 and BRB~21 \citep{casewell2007}, MKO
$K$~band of HHJ~6, Roque~4, CFHT~17, and Roque~7 \citep{pinfield2003}, and
$J$~band of MHObd~1 (2MASS, converted to UKIDSS photometric system).}\\
\tablefoottext{c}{References for the quadruplet (SpT,Li,$\mu$,$I_{\rm C}$):\\
{\sl Spectral type or lithium test}: 1) \citet{steele1995a}, 2)
\citet{basri1996}, 3) \citet{rebolo1996}, 4) \citet{martin1996}, 5)
\citet{cossburn1997}, 6) \citet{zapateroosorio1997b}, 7) \citet{martin1998a}, 8)
\citet{stauffer1998a}, 9) \citet{festin1998b}, 10) \citet{stauffer1998b}, 11)
\citet{martin1998b}, 12) \citet{martin2000}, 13) \citet{pinfield2003}, 14) this
study. Spectral type uncertainties for our L-type targets are listed
in Table~\ref{log}; for the other objects the uncertainties are typically of half
a subclass.\\
{\sl Proper motion}: 1) \citet{hambly1993}, 2) \citet{pinfield2000}, 3)
\citet{moraux2001}, 4) \citet{deacon2004}, 5) \citet{bihain2006}, 
6)~\citet{casewell2007}, 7) \citet{lodieu2007}, 8) \citet{stauffer2007}, 9) this study.\\
{\sl $I_{\rm C}$-band photometry}: 1) \citet{stauffer1994}, 2)
\citet{festin1998a}, 3) \citet{bouvier1998}, 4) \citet{stauffer1998b}, 5)
\citet{jameson2002}, 6)~\citet{bihain2006}, 7) \citet{stauffer2007}. Magnitude
errors are of 0.05-0.1~mag.}\\
\tablefoottext{d}{Roque~4 has radial velocity and spectral features consistent with cluster membership \citep{zapateroosorio1997b,kirkpatrick2008}.}
\end{flushleft}
      }
\normalsize
\end{table*}

\section{Cluster membership confirmation of Roque~25 by proper
motion}\label{pm25}

As part of the project to characterize the Pleiades L-type population by both
proper motion and spectroscopy, we obtained  $H$-band imaging data of
\object{Roque~25}, the first L-type object identified in the Pleiades
\citep{martin1998b,martin2000}, to assess its cluster membership by proper
motion. The observations were done on the night of 2006 February 15 with the
Omega-2000 instrument ($15.4\times15.4$~arcmin$^2$, 0.45~arcsec~${\rm
pix}^{-1}$) mounted at the 3.5~m Telescope (Centro Astron{\'o}mico
Hispano-Alem{\'a}n de Calar Alto, Spain). They consisted of 1.6~s
$\times$~23~coadds $\times$~36~dithers on the science target, as well as dome
flats. These data were reduced as in \citet{bihain2006}. We performed aperture
and point-spread-function (PSF) photometry using routines within the {\tt
DAOPHOT} package. The $H$-band photometry was calibrated using 341 stellar
sources from UKIDSS GCS DDR6, with aperture corrected magnitudes of errors
$\sigma_{H~(\rm UKIDSS)}<0.1$~mag (2.0~arcsec aperture diameter). The photometry
has an average relative calibration error of 0.04~mag. For Roque~25, we measure
$H=16.85\pm0.05$~mag, in agreement with $H {\rm (UKIDSS)}=16.86\pm0.04$~mag.
Using the Omega-2000 image of Roque~25 and the discovery $I$-band image
\citep{zapateroosorio1999Pl}, corresponding to a time baseline of 10.01~yr, and
the method described in \citet{bihain2006}, we obtain a proper motion $(\mu_{\rm
\alpha} {\rm cos} \delta,~\mu_{\rm \delta})$
=~(27.4~$\pm$~7.3,~$-$32.6~$\pm$~3.9)~mas~yr$^{-1}$. Considering the average
cluster proper motion $(\mu_{\rm \alpha} {\rm cos} \delta,~\mu_{\rm \delta})$
=~(19.15~$\pm$~0.23,~$-$45.72~$\pm$~0.18)~mas~yr$^{-1}$ \citep{robichon1999} and
the fact that the low mass cluster members have a larger intrinsic velocity
dispersion than that of the high mass members \citep{pinfield1998,bihain2006},
our result confirms that Roque~25 is also a Pleiades member by proper motion.


\end{document}